\documentclass[12pt,preprint]{aastex}
\usepackage{psfig}
\usepackage{graphicx}

\shorttitle{HR 9024 X-ray flare}
\shortauthors{Testa et al.}
\usepackage{color} 
\usepackage{amsmath}           
\usepackage{amsfonts}

\def \lya  {Ly$\alpha$}

\def \lx   {$L_{\rm X}$}

\def \hr       {HR~9024}
\def \cha      {{\em Chandra}}
\def \xmm      {XMM-{\em Newton}}
\def \hetgs    {{\sc hetgs}}
\def \hetg     {{\sc hetg}}

\def \fexvii  {Fe\,{\sc xvii}}
\def \fexviii {Fe\,{\sc xviii}}

\def \fexxi   {Fe\,{\sc xxi}}

\def \fexxv   {Fe\,{\sc xxv}}

\def \neix   {Ne\,{\sc ix}}
\def \nex    {Ne\,{\sc x}}
\def \ovii   {O\,{\sc vii}}
\def \oviii  {O\,{\sc viii}}
\def \mgxi   {Mg\,{\sc xi}}
\def \mgxii  {Mg\,{\sc xii}}
\def \sixiii {Si\,{\sc xiii}}
\def \sxv    {S\,{\sc xv}}
\def \sxvi   {S\,{\sc xvi}}
\def \sixiv  {Si\,{\sc xiv}}
\def \caxix  {Ca\,{\sc xix}}
\def \caxx   {Ca\,{\sc xx}}
\def \sxvi   {S\,{\sc xvi}}
\def \sxv    {S\,{\sc xv}}
\def \nvii   {N\,{\sc vii}}
\def \arxvii  {Ar\,{\sc xvii}}
\def \arxviii {Ar\,{\sc xviii}}
\def \alxiii  {Al\,{\sc xiii}}
\def \alxii   {Al\,{\sc xii}}

\def \nvii    {N\,{\sc vii}}

\begin{document}
\title{Detailed diagnostics of an X-ray flare in the single giant HR 9024}
\author{Paola Testa\altaffilmark{1}, Fabio Reale\altaffilmark{2,3}, 
David Garcia-Alvarez\altaffilmark{4}, David P.\ Huenemoerder\altaffilmark{1}}
\altaffiltext{1}{Massachusetts Institute of Technology, Kavli 
Institute for Astrophysics and Space Research, 70 Vassar street, 
Cambridge, MA 02139, USA; testa@space.mit.edu}
\altaffiltext{2}{Dipartimento di Scienze Fisiche \& Astronomiche, 
Sezione di Astronomia, Universit\`a di Palermo Piazza del Parlamento 
1, 90134 Palermo, Italy}
\altaffiltext{3}{INAF - Osservatorio Astronomico di Palermo,
Piazza del Parlamento 1, 90134 Palermo, Italy}
\altaffiltext{4}{Imperial College London, Blackett Laboratory,
Prince Consort Rd, London, SW7 2AZ, UK}

\begin{abstract}
\end{abstract}

We analyze a 96~ks \cha\ \hetgs\ observation of the single G-type 
giant \hr. The high flux allows us to examine spectral line and 
continuum diagnostics at high temporal resolution, to derive plasma 
parameters.
A time-dependent 1D hydrodynamic model of a loop with half-length 
$L = 5 \times 10^{11}$~cm ($\sim R_{\star}/2$), cross-section radius
$r = 4.3 \times 10^{10}$~cm, with a heat pulse of 15~ks and 
$2 \times 10^{11}$~erg~cm$^{-2}$~s$^{-1}$ deposited at the loop 
footpoints, satisfactorily reproduces the observed evolution of 
temperature and emission measure, derived from the analysis of the 
strong continuum emission. 
For the first time we can compare predictions from the hydrodynamic 
model with single spectral features, other than with global spectral 
properties.
We find that the model closely matches the observed line emission, 
especially for the hot ($\sim 10^8$~K) plasma emission of the \fexxv\ 
complex at $\sim 1.85$\AA. 

The model loop has $L/R_{\star} \sim 1/2$ and aspect ratio 
$r/L \sim 0.1$ as typically derived for flares observed in active 
stellar coronae, suggesting that the underlying physics is the same 
for these very dynamic and extreme phenomena in stellar coronae 
independently on stellar parameters and evolutionary stage.


\section{Introduction}
\label{s:intro}

The first X-ray stellar surveys showed widespread presence of coronal 
emission in the cool half of the H-R diagram (see e.g., 
\citealt{Vaiana81} for a review).
Improved spatial and spectral resolution have provided us with more 
powerful tools for investigating the characteristics of the X-ray 
coronal activity in late-type stars, and for exploring the underlying 
processes at work in stars with different stellar parameters, and 
evolutionary stage.

X-ray observations of late-type stars show that coronal phenomena 
observed at close range on the Sun are common in late-type stars 
though they occur on more extreme scales in very active stars. 
For instance, stars at higher activity levels have reach much higher 
temperatures and densities (e.g., \citealt{Sanz02,Testa04b,Ness04})
than typically observed for solar coronal plasma. Analogously, coronal
flares are routinely observed in late-type stars with characteristics
similar to those observed in solar flares, e.g., fast rise and slow 
decay (e.g., \citealt{Reale02} for a review), but at the same time the
flare frequency and intensity can be dramatically larger than observed 
on the Sun, with stellar X-ray luminosity increasing orders of 
magnitude with respect to the quiescent level (e.g., 
\citealt{Favata00,Osten06}).

The analysis of lightcurves during flares provides us with insights 
into the characteristics of the coronal structures and therefore of 
the magnetic field (e.g., \citealt{Schmitt99,Favata00,Reale04}). 
Even though stellar flares are spatially unresolved, a great deal of 
information on the coronal heating and on the plasma structure 
morphology can be inferred from the detailed modeling of stellar 
flares; for instance, if enough data statistics is available for
moderately time-resolved spectral analysis, the study of the complete 
evolution of a flare allows to infer whether the flare occurs in 
closed coronal structures (loops), what is the size of these flaring 
structures, whether continuous heating is present throughout the 
flare, and even constraints on the location and distribution of the 
heating (see \citealt{Reale04}, and \S\ref{s:res_modeling} for a 
detailed discussion).

This work presents a detailed modeling of a large X-ray flare on the 
single evolved G1~III giant \hr, observed with the \cha\ High Energy 
Transmission Grating.
Most of the flare is observed, from the rise phase to the late decay, 
allowing to constrain a detailed hydrodynamic simulation of the 
flaring structure.
The analysis of this flare is especially interesting in the context of
X-ray activity of evolved giants. \hr\ is an intermediate mass star 
($M_{\star} \sim 2.9 M_{\odot}$) in the Hertzsprung gap, that is in 
its initial rapid ($< 1$~Myr) phase of post-main sequence evolution 
when it enters the cool region of the H-R diagram, and it develops a 
subphotospheric convective layer (see e.g., \citealt{Pizzolato00}). 
The intermediate mass Hertzsprung gap giants are strong X-ray emitters 
($L_{\rm X} \sim 10^{31}$~erg~s$^{-1}$) while their main-sequence 
progenitors, late-B or early-A type, are X-ray dark lacking the 
fundamental ingredients to sustain X-ray activity, either 
magnetic-dynamo as in late-type stars or strong winds as in massive 
stars. 
The onset of an efficient convective layer, together with their 
typically fast rotation rates (due to the little if any loss of 
angular momentum in their main sequence phase) is thought to generate 
a dynamo mechanism sustaining the X-ray activity of these evolved 
yellow giants.
The young coronae of these stars are characterized by high temperature
and density (e.g., \citealt{Ayres98,Testa04b}), similarly to low-mass 
active stars at the same high activity levels (e.g., 
\citealt{Sanz02,Testa04b}); on the other hand, there is evidence for 
significant differences such as much lower coronal filling factors 
\citep{Testa04b}, and very limited flaring activity.
The evolved intermediate mass giants, both in the Hertzsprung gap and 
in the post helium flash ''clump" (the relatively long-lived, 
$\sim 70$~Myr, core Helium burning phase; \citealt{Ayres99}), show 
extremely constant X-ray emission level (e.g., 
\citealt{Haisch94,Gondoin03,Testa04b,Audard04,Scelsi04}) and only a few 
flares have been observed on these sources (\citealt{Haisch94,Ayres99}, 
and this work).

The opportunity of deriving a loop length from the analysis of the 
flare is interesting also in that it gives us an opportunity to probe 
the structuring of the corona, which in principle can be significantly 
different from dwarf stars coronae. In fact in these giants the 
gravity is considerably lower than for MS late-type stars therefore 
yielding a larger scale height, and possibly allowing the existence of 
very extended coronae (as suggested for example by \citealt{Ayres03}). 
In the case of \hr\ the surface gravity is only $\sim 0.02 g_{\odot}$ 
implying a scale height of about 3 stellar radii at $10^7$~K 
(30$R_\star$ at $10^8$~K).

\hr\ (HD~223460, OU~And) is a moderately rotating 
($v\sin{i} \sim 21$~km~s$^{-1}$, \citealt{DeMedeiros92}) 
chromospherically active single giant, not too well studied even 
though it is a bright and close by object ($d \sim 135$~pc, 
\citealt{Perryman97}).
The stellar parameters are listed in Table~\ref{tab:param_star}. 
Several X-ray observations of \hr\ exist indicating high and constant
X-ray luminosity of a few $10^{31}$~erg~s$^{-1}$. 
\cite{Singh96a} analyzed ROSAT {\sc pspc} observations of 
chromospherically active stars and derived an X-ray luminosity 
\lx$\sim 4.3 \times 10^{31}$~erg~s$^{-1}$ in the $0.2-2.4$~keV energy 
band; this translates to \lx$\sim 2.6 \times 10^{31}$~erg~s$^{-1}$ 
using the revised distance of 135~pc instead of 175~pc assumed by 
\cite{Singh96a}. This value is in a good agreement with values 
obtained by \cite{Gondoin03}, in the $0.3-2$~keV range, from two short 
\xmm\ observations of \hr\ showing little if any variability: 
\lx$\sim 2.7 \times 10^{31}$~erg~s$^{-1}$ (Rev.107, 6~ks exposure 
time), and \lx$\sim 2.3 \times 10^{31}$~erg~s$^{-1}$ (Rev.200, 
3~ks exposure time); the total X-ray luminosities in the $0.3-10$~keV 
range are $3.8 \times 10^{31}$~erg~s$^{-1}$ and 
$3.0 \times 10^{31}$~erg~s$^{-1}$ respectively.

\begin{table}[!hb]
  \begin{center}
    \caption{Stellar parameters.}
    \begin{tabular}[h]{cccccc}
      \hline
  &  &  &  &  &    \\ 
 Spec.\ type  &  d  &  $R$  &  $M$  &  $\log L_{\rm bol}$  &  $P_{\rm rot}$\\[-0.1cm]
    & [pc] & [$R_{\odot}$] & [$M_{\odot}$] &  [erg~s$^{-1}$]  &  [days] \\  \hline
  G1 {\sc iii} & 135\tablenotemark{a} & 13.6\tablenotemark{b} & 2.9\tablenotemark{c} & 
  35.4\tablenotemark{d}  & 23.25\tablenotemark{e} \\ \hline \\
      \vspace{-2cm}
 \tablenotetext{}{References:(a) from SIMBAD; (b) \cite{Singh96a}; 
 	(c) \cite{Gondoin99}; (d) \cite{Flower96}; (e) \cite{Singh96a}.}
      \end{tabular}
    \label{tab:param_star}
  \end{center}
\end{table}

Our approach here is to inspect the flare lightcurve and spectra and 
derive some quantities relevant to set up a loop hydrodynamic model 
and to constrain the initial parameters.
The numerical solution of the hydrodynamic equations allow us to
synthesize in detail the emission (the so-called {\it forward modeling})
as it would be really observed and therefore to compare directly to the
data. This allows us to have a feedback on the model and to refine the 
model parameters. Once obtained a good description of the global 
features, i.e. the lightcurve and the overall evolution of the 
temperature and total emission measure, the model will have constrained 
the loop length and aspect, and the heating function.
The comparison with the fine details of the data analysis will give us 
insight into the flare density and thermal structure and evolution, and 
help us to interpret the results of the data analysis.

The paper is structured as follows: we describe the observations in 
\S\ref{s:obs}, and the methods both for the spectral analysis and the 
hydrodynamic modeling in \S\ref{s:analysis}; in \S\ref{s:results} we 
present the results of our study, and in \S\ref{s:discuss} we draw our 
conclusions.

\section{Observations}
\label{s:obs}

We analyzed the \cha\ High Energy Transmission Grating 
Spectrometer (see \citealt{Canizares00,hetg05} for a description of 
the instrumentation) observation of the X-ray active single G1 giant
HR~9024.  
The data were obtained from the \cha\ Data 
Archive\footnote{http://cxc.harvard.edu/cda} and have been reprocessed 
using standard CIAO 3.3 tools and analysis threads.  
Effective areas and line responses (ARFs and RMFs) were calculated using 
standard CIAO procedures\footnote{see http://cxc.harvard.edu/ciao/threads/}.  
The characteristics of the \hetg\ observation are listed in
Table~\ref{tab:param_obs}. 

\begin{table}[!hb]
  \begin{center}
    \caption{Parameters of \hetg\ observation.}
    \begin{tabular}[h]{cccccc}
      \hline
  ObsID  &  Start  &  End  &  $t_{\rm exp}$  &  
  $\log L_{\rm X}$\tablenotemark{a}  &  $\log L_{{\rm X},peak}$\tablenotemark{b} \\[-0.1cm]
         &         &       &  [ks]   & [erg~s$^{-1}$]  & [erg~s$^{-1}$] \\
      \hline
         & 2001-08-11  &  2001-08-12  &      &       &       \\[-0.2cm]
  1892   &  00:19:04   &   03:43:14   & 95.7 & 31.8  &  32.1 \\
      \hline \\
      \vspace{-2cm}
 \tablenotetext{a}{From {\sc meg} spectrum, in the 2-24\AA\ range. 
 	Averaged over the whole observation.}
 \tablenotetext{a}{Peak X-ray luminosity obtained from the {\sc meg} 
 	spectrum integrated in the time interval 10-15~ks from 
	the beginning of the observation.}
      \end{tabular}
    \label{tab:param_obs}
  \end{center}
\end{table}

\begin{figure}[!h]
\centerline{\psfig{figure=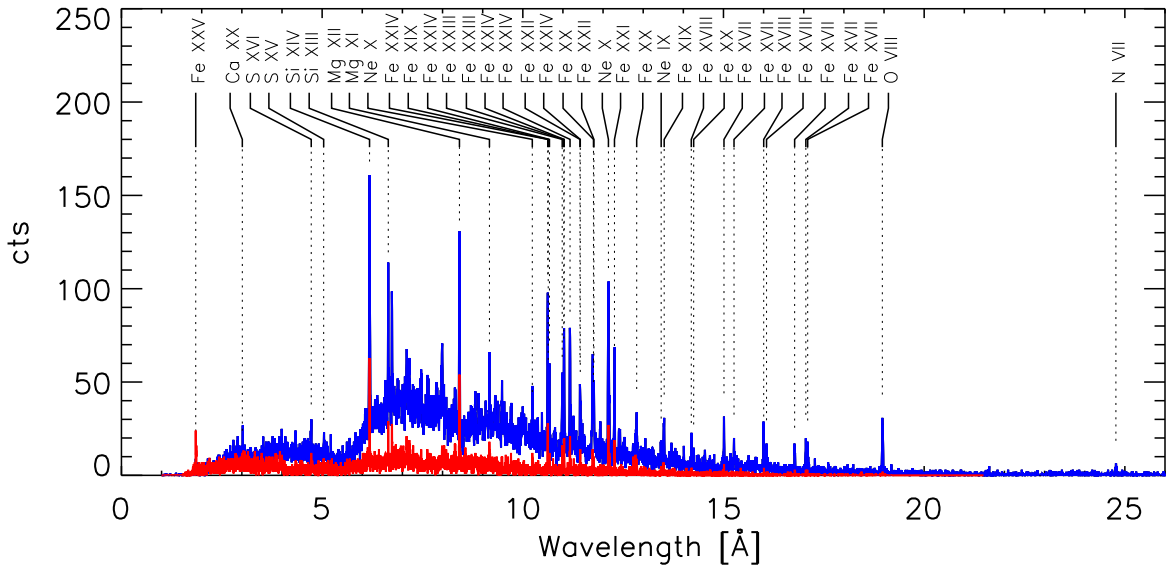,width=17cm}}
\vspace{-0.5cm}
\caption{\cha\ {\sc heg} (red) and {\sc meg} (blue) spectra obtained
	in a 96~ks observation of the single giant HR~9024.
	Line identification for many prominent spectral features is 
	provided.
 \label{fig:spectrum}}
\end{figure}

Figure~\ref{fig:spectrum} shows the \cha\ High Energy ({\sc heg}) and 
Medium Energy ({\sc meg}) Gratings spectra for the $\sim 96$~ks 
observation of \hr. 
The spectrum is characterized by a strong continuum emission, 
indicating high temperature plasma, as well as the unusually strong 
\fexxv\ complex at $\sim 1.85$\AA\ ($\log T[$K$] \sim 7.8$), and \caxx\ 
resonance line at $3.02$\AA\ ($\log T[$K$] \sim 7.7$); these 
characteristics are somewhat extreme when compared with the spectra of 
other very active stellar coronae
(see e.g.,\citealt{Testa04b}). Other prominent features in the spectrum 
are the H-like and He-like lines of Si ($\sim 6.2, 6.7$\AA\ 
respectively) and Mg ($\sim 8.4, 9.2$\AA), and the strong emission lines
of highly ionized Fe around 11\AA.

\begin{figure}[!h]
\centerline{\hspace{-0.5cm}
\psfig{figure=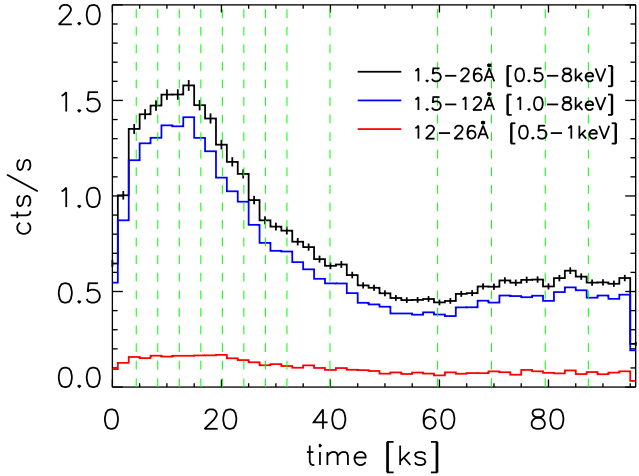,width=10cm}}
\vspace{-0.5cm}
\caption{Lightcurve obtained as the sum of total counts of {\sc heg} 
	and {\sc meg} dispersed spectra, using a temporal binsize of 
	2~ks. Lightcurves in hard (blue) and soft (red) spectral bands 
	are shown. The vertical green dashed lines mark the time 
	intervals selected for the temporally resolved spectral 
	analysis and the hydrodynamic modeling (see \S\ref{s:analysis}).
	\label{fig:lightcurve}}
\end{figure}

The lightcurve of the summed {\sc heg}+{\sc meg} dispersed photons 
integrated in the 1.5-26\AA\ range, presented in 
Figure~\ref{fig:lightcurve}, shows clear variability: the X-ray 
emission level rises steeply at the beginning of the observation by a
factor $\sim 3$ in about 15~ks; after a slow decay on a timescale of 
about 40~ks the lightcurve rises again, with peak around 80~ks from
the beginning of the observation. 
The lightcurves for a hard (1.5-12\AA) and a soft (12-26\AA) spectral 
band, also shown in Fig.~\ref{fig:lightcurve}, indicate hardening of the
spectrum corresponding to the two peaks of the emission, typical of 
stellar flares.
The peak luminosity above $10^{32}$~erg~s$^{-1}$ is extremely high when
compared with typical energies of stellar flares observed in active coronae
of late-type stars.

\section{Analysis}
\label{s:analysis}

Spectra were analyzed with the 
PINTofALE\footnote{http://hea-www.harvard.edu/PINTofALE}
IDL\footnote{Interactive Data Language, Research Systems Inc.}
software \citep{PoA}. 
The high resolution spectra provide several plasma diagnostics 
(temperature, density, abundances, emission measure distribution),
from the analysis of both continuum and emission lines, and from the 
lightcurves in different spectral bands or in single lines.
The high flux allows us to examine spectral line and continuum 
diagnostics at high temporal resolution.

\subsection{Spectral analysis}
\label{s:analysis_contlines}
\paragraph{Continuum emission:}
The strong continuum emission in this spectrum provides us with 
constraints on the hot plasma on which it strongly depends. 
Specifically, temperature and emission measure, EM, can be estimated 
through a fit of the continuum, and their evolution, probing the 
hottest plasma component, allows us to constrain the hydrodynamic 
model of the flaring structures (see \S\ref{s:analysis_hd}, and 
\S\ref{s:res_modeling}). 
In order to derive T and EM, we fit the continuum simultaneously in 
the {\sc heg} and {\sc meg} spectrum, selecting narrow spectral 
regions which can be assumed as "line-free" to a reasonable extent, 
on the basis of predictions of the atomic databases APED 
\citep{Smith01}, and CHIANTI \citep{chianti,chianti06}.
We fit the continuum with an isothermal model computed in PINTofALE, 
using  CHIANTI, and convolved with the \hetg\ spectral response. 
This model contains all the contributions to the continuum (free-free, 
free-bound, 2 photon), however at high temperature 
($\log T[$K$] \gtrsim 7.5$), as observed in this spectrum, the 
bremsstrahlung continuum is by far the dominant process for the 
formation of the continuum.
The "line-free" spectral regions used for the fit are: $2.00-2.95$\AA,
$4.4-4.6$\AA, $5.3-6.0$\AA, $7.5-7.8$\AA, $12.5-12.7$\AA, $19.1-20$\AA.
The fit also provides an estimate for EM from the normalization 
parameter.
The analysis of the strong continuum emission allows us to derive the 
evolution of the temperature and emission measure during the flare with 
high temporal resolution. The time intervals chosen for this analysis 
are shown in Fig.~\ref{fig:lightcurve} superimposed on the lightcurve; 
they have been selected in order to have high temporal resolution but 
also good constraints on the plasma parameters.

\paragraph{Line emission:}
The line fluxes are determined using the technique of spectral fitting 
described in \cite{Testa04b}.  
The measured line fluxes have been used for the reconstruction of the 
emission measure distribution and the determination of abundances 
(\S\ref{s:res_demabund}), and for the comparison with the results of 
the hydrodynamic model (\S\ref{s:res_modeling}).

\paragraph{Abundances and Emission Measure Distribution:} 
The emission measure distribution, EM(T), is derived through a 
Markov-Chain Monte-Carlo analysis using the Metropolis algorithm 
(MCMC[M]; \citealt{MCMC98}) on a set of line flux ratios, as in 
\cite{GarciaA06}. O lines are the coolest lines used and Ar lines 
the hottest, covering a temperature range $\log T[$K$] \sim 6.2-7.8$. 
Coronal abundances are evaluated on the basis of the derived EM(T): 
the abundance is a scaling factor in the line flux equation to match 
the measured flux \citep{GarciaA06}.

\subsection{Hydrodynamic modeling}
\label{s:analysis_hd}
The inspection of the light curve and of the evolution of the 
temperature and of the integrated emission measure (EM) during the 
flare allows to set up a detailed model of the flaring structure(s) 
\citep{Reale04}, as described in detail in \S\ref{s:res_modeling}. 
We derive the temperature and EM parameters through the analysis of the 
continuum emission as described above (\S\ref{s:analysis_contlines}), 
the continuum emission being strong and probing the hottest plasma.  

The 1D hydrodynamic model, solves time-dependent plasma equations with 
detailed energy balance \citep{Peres82,Betta97}, with a time-dependent 
heating function defining the energy release triggering the flare (see 
e.g., \citealt{Reale97,Reale04}). 
The coronal plasma is confined in a closed loop structure, where plasma 
motion and energy transport occur only along magnetic field lines. 
For the initial atmosphere we assume a loop in hydrostatic equilibrium
and detailed energy balance \citep{S81} with maximum temperature (at the 
apex) $T_{\rm max}=2 \times 10^7$~K; the initial conditions do not 
influence much the evolution of the plasma after a very short time.

The outputs of the hydrodynamic simulations are distributions of 
temperature and density along the loop sampled at regular times 
throughout the flare evolution. 
From these plasma parameters we synthesize the corresponding \hetg\ 
spectrum of each (isothermal) plasma volume along the loop at each 
given time, using isothermal models folded with the \hetg\ spectral 
response. We then integrate along the loop to obtain the overall 
\hetg\ spectrum of the multi-thermal plasma in the flaring structure.

\section{Results}
\label{s:results}
\subsection{Emission measure distribution and abundances}
\label{s:res_demabund}
As a first step we carry out an analysis of the spectrum to study the 
global characteristics of thermal distribution and abundances.
To obtain this information we need strong lines, from different 
elements, covering a wide temperature range. We have to integrate on 
time intervals long enough to have good photon statistics for each 
line.  This limits our temporal resolution and we are not able to 
derive EM(T) and abundances in several portions of the flare, but only 
in two different portions of the observation: during the flare (i.e.\ 
using the spectrum integrated over the first 40~ks of the observation), 
and outside the flare (40-96~ks). We note that hereafter we will label 
the parameters derived outside the flare as ''quiescent''; even though 
the corona seems to undergo another dynamic event this second flare is 
on a smaller scale with respect to the first one.
The analysis of the abundances is useful also for the synthesis of the 
spectra from the results of hydrodynamic modeling and a consistent 
comparison with the data.

Table~\ref{tab:fluxes} lists the fluxes of the spectral lines used for 
this analysis, measured in the two phases of flare and quiescence.

\begin{deluxetable}{llcrrc}
\tablecolumns{6} 
\tabletypesize{\footnotesize}
\tablecaption{Measured fluxes (in $10^{-6}$~photons~cm$^{-2}$~sec$^{-1}$)
	of the spectral lines used in our analysis, during the flare and 
	outside the flare.
	\label{tab:fluxes}}
\tablewidth{0pt}
\tablehead{
 \colhead{ion}                                     &  \colhead{$\lambda_{\rm obs}$ [\AA]}  &  
 \colhead{$\log T_{\rm max}$[K]\tablenotemark{a}}  &  \colhead{flux$_{\rm flare}$}         &  
 \colhead{flux$_{\rm quiesc}$}                     &  \colhead{Use\tablenotemark{b}} 
}
\startdata 
\fexxv\      	  &  1.853    & 7.8 &   94   $\pm$ 19   &    13.5  $\pm$ 8.4 &   M \\
\fexxv\      	  &  1.864    & 7.8 &   49   $\pm$ 16   &    16.0  $\pm$ 8.4 &   M \\
\caxx\       	  &  3.024    & 7.7 &   21.7 $\pm$ 6.8  &    $<4.4$	     &   S \\
\caxix\      	  &  3.187    & 7.4 &   18.7 $\pm$ 6.4  &    5.8   $\pm$ 3.7 &   S \\
\arxviii\    	  &  3.734    & 7.6 &   14.1 $\pm$ 7.0  &    4.2   $\pm$ 3.6 &   S \\
\arxvii\     	  &  3.945    & 7.3 &   13.5 $\pm$ 6.7  &    4.7   $\pm$ 4.0 &   S \\ 
\sxvi\       	  &  4.729    & 7.4 &   20.3 $\pm$ 8.7  &    10.8  $\pm$ 5.7 &   S \\
\sxv\        	  &  5.041    & 7.2 &   30   $\pm$ 11   &    9.5   $\pm$ 6.0 &   S \\
\sixiv\      	  &  6.183    & 7.2 &   82.9 $\pm$ 6.0  &    42.2  $\pm$ 3.6 &  SM \\
\sixiii\     	  &  6.648    & 7.0 &   33.1 $\pm$ 4.2  &    25.7  $\pm$ 3.0 &   S \\
\alxiii\          &  7.170    & 7.1 &   15.9 $\pm$ 4.5  &    8.1   $\pm$ 3.1 &   S \\
\alxii\           &  7.759    & 6.9 &   6.2  $\pm$ 3.8  &    5.6   $\pm$ 3.1 &   S \\
\mgxii\           &  8.422    & 7.0 &   80.7 $\pm$ 6.4  &    47.5  $\pm$ 4.5 &  SM \\  
\mgxi\            &  9.168    & 6.8 &   31.0 $\pm$ 4.7  &    15.2  $\pm$ 2.8 &   S \\
\nex\             &  12.132   & 6.8 &   142. $\pm$ 12.  &    113.  $\pm$ 9.3 &   S \\
\fexxi\           &  12.284   & 7.0 &   79.  $\pm$ 11.  &    53.5  $\pm$ 7.2 &   S \\
\neix\            &  13.448   & 6.6 &   20.  $\pm$ 10.  &    22.6  $\pm$ 7.4 &   S \\  %
\fexviii\         &  14.201   & 6.9 &   50.  $\pm$ 15.  &    30.1  $\pm$ 9.1 &   S \\  %
\fexvii\          &  15.012   & 6.7 &   107. $\pm$ 22.  &    43.   $\pm$ 13. &   S \\  %
\oviii\           &  18.965   & 6.5 &   270. $\pm$ 42.  &    179.  $\pm$ 29. &   S \\  %
\ovii\            &  21.602   & 6.3 &   51   $\pm$ 44.  &    $<25$ 	     &   S \\  %
\nvii\            &  24.778   & 6.3 &   65.  $\pm$ 50.  &    72.   $\pm$ 40. &   S \\ %
 \enddata
\tablenotetext{a}{Temperature of maximum formation of the line.}
\tablenotetext{b}{Use indicates whether the line was used in the spectral analysis ("S"),
	for the emission measure reconstruction and abundance determination, and whether 
	the feature was used for the direct comparison with the loop model ("M"; see 
	\S\ref{s:res_modeling}).}
\end{deluxetable}

\begin{figure}[!h]
\centerline{\psfig{figure=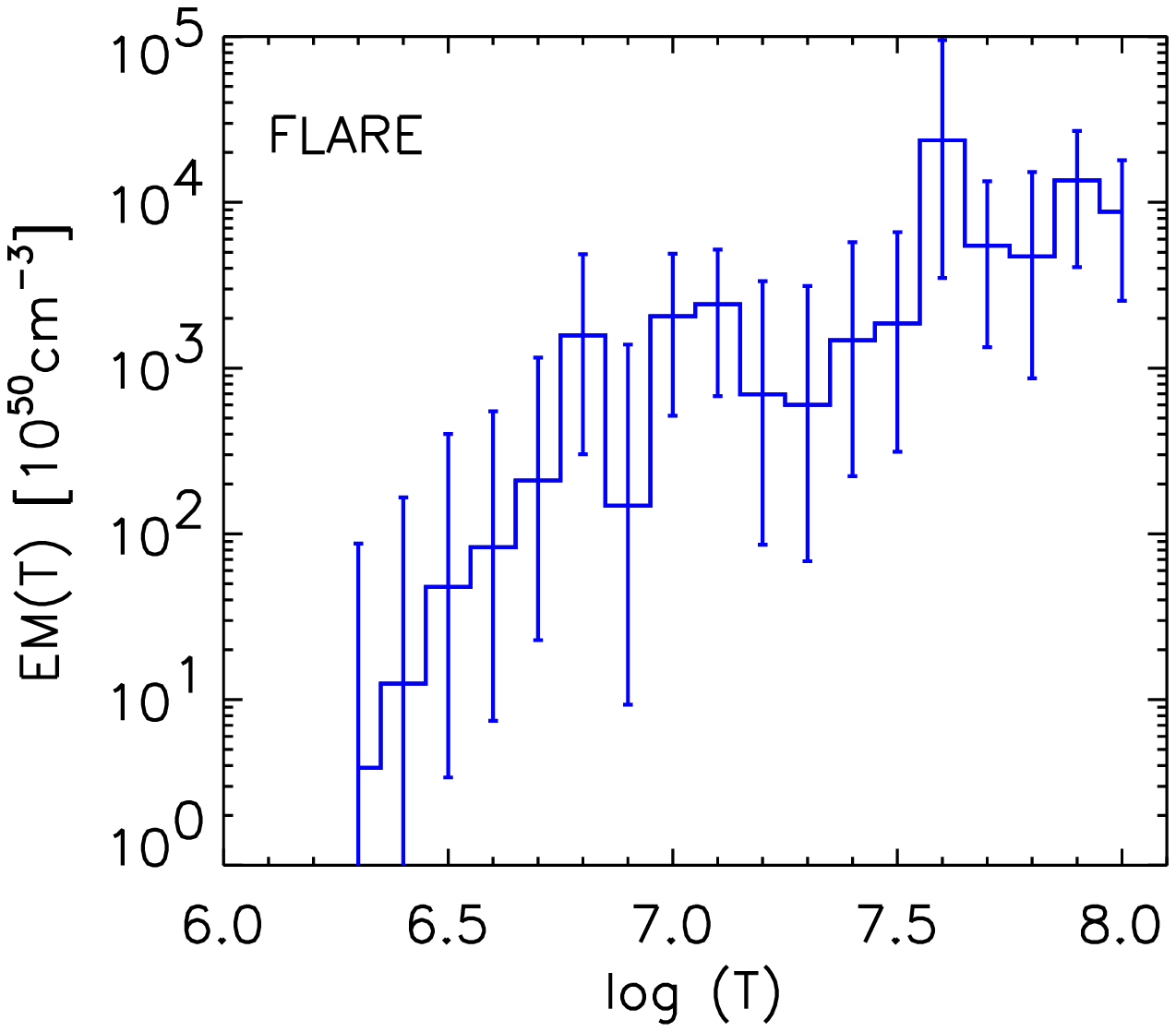,width=8cm}\hspace{-0.6cm}
	    \psfig{figure=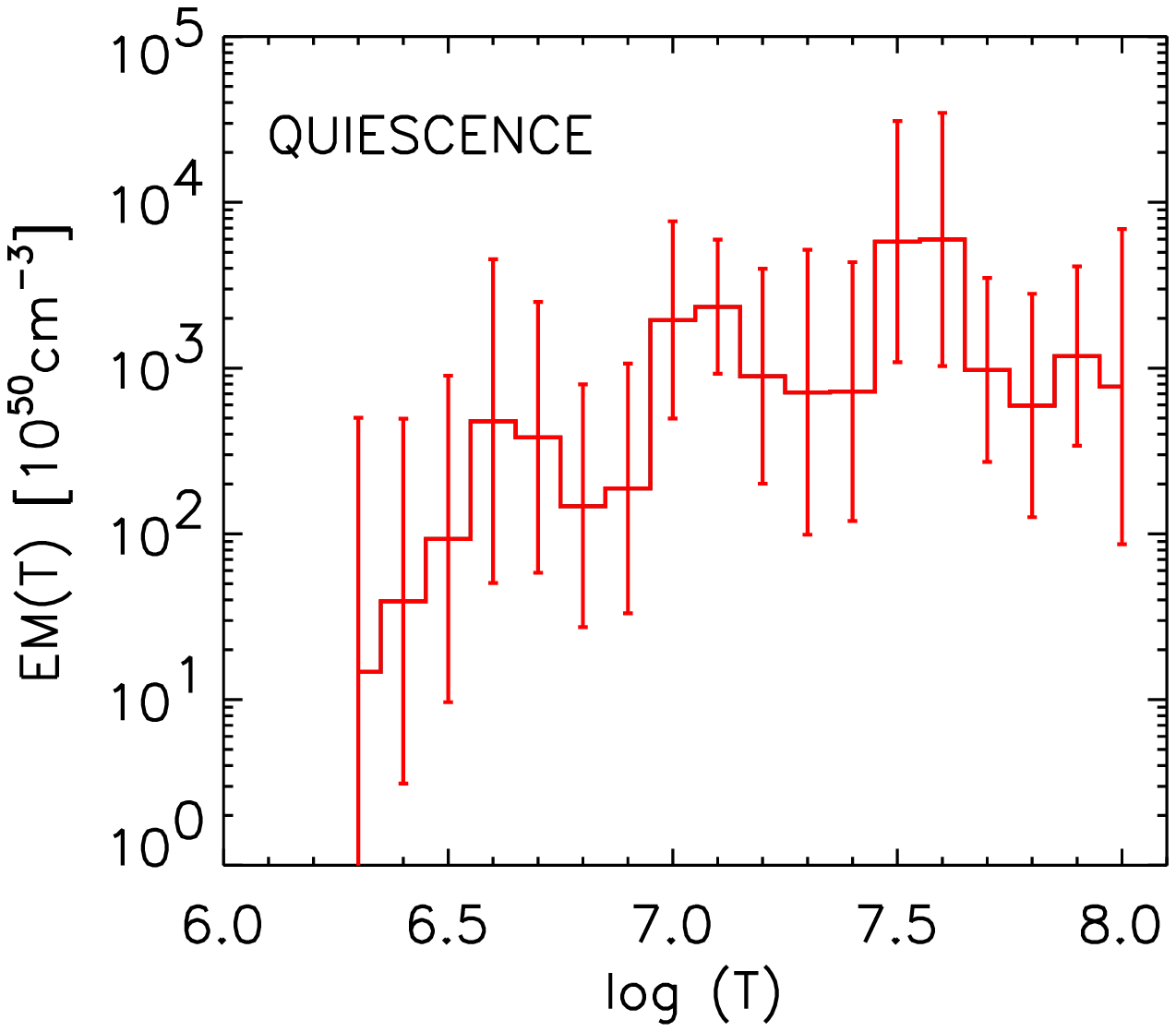,width=8cm}}
\vspace{-0.3cm}
\caption{Emission measure distribution derived from the flare portion
	of the spectrum (0-40~ks; {\em left}), and from the spectrum 
	outside the flare (40-96~ks; {\em right}). The comparison 
	shows that the main difference between flare and quiescent 
	emission resides in the hot end of the temperature range, 
	i.e.\ for $\log T[$K$] \gtrsim 7.5$ where the EM(T) of the 
	flaring plasma is about one order of magnitude higher than 
	the EM(T) of the plasma outside the flare.
	\label{fig:dem}}
\end{figure}

\begin{figure}[!h]
\centerline{\psfig{figure=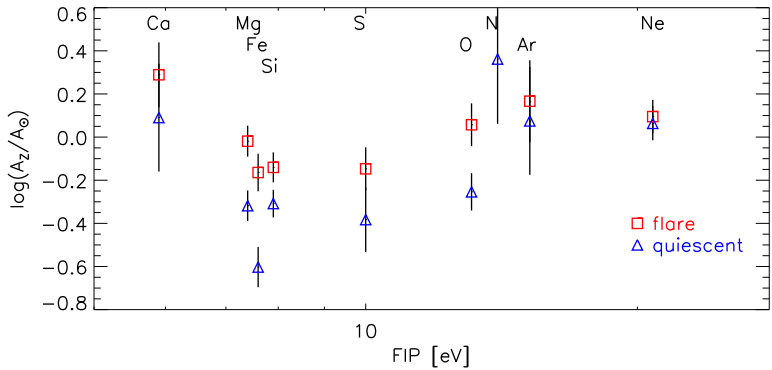,width=11cm}}
\vspace{-0.3cm}
\caption{Coronal abundances relative to solar abundances 
	\citep{Asplund05} as a function of the First Ionization 
	Potential (FIP) derived for \hr\ during the flare (square 
	symbols) and the quiescent (triangles) phase.
	\label{fig:abund}}
\end{figure}

Fig.~\ref{fig:dem} and \ref{fig:abund} show the emission measure 
distribution and the coronal abundances as derived from the flare 
spectrum, and from the quiescent emission.

The thermal distribution of the coronal plasma in \hr\ appears 
dominated by hot (i.e., typical of flaring structures) plasma both 
during and outside the flare. Hot emission ($T \gtrsim 2 \times 10^7$~K)
was found also from the analysis of two very short (6~ks, and 3~ks) 
\xmm\ observations of \hr\ when the corona seems to be in its 
quiescent state \citep{Gondoin03}.
The main difference between the EM(T) of flare and quiescence is at the
hot end of the temperature range, i.e.\ for $\log T[$K$] \gtrsim 7.5$,
where the EM(T) of the flaring plasma is about one order of magnitude 
higher than the EM(T) of the plasma outside the flare.
The EM(T) derived for \hr\ presents interesting characteristics when 
compared to other active coronae. Specifically, the EM(T) of \hr\ is 
characterized by a rather shallow slope, similar to the slope of 
hydrostatic loop models ($EM(T) \propto T^{3/2}$; \citealt{RTV}), 
whereas for several active coronae there is increasing evidence of 
steep, almost isothermal, emission measure distributions possibly 
indicating the dynamic nature of the coronal loops composing them 
\citep{Testa05a}: for instance, such steep EM(T) are derived for 
other giants, e.g., the Hertzsprung gap giant 31~Com \citep{Scelsi04}, 
and the clump giants $\beta$~Cet, \citep{Sanz02}, and Capella 
\citep{Dupree93}, and for other active stars (e.g., 
\citealt{Griffiths98,Drake00,Sanz02}).

The coronal abundance pattern observed in \hr\ (Fig.~\ref{fig:abund}) 
presents characteristics similar to other intermediate mass giants,
i.e.\ little or no FIP effect \citep{GarciaA06}. 
All abundances are plotted relative to the solar mixture of 
\cite{Asplund05} as photospheric abundances for \hr\ have not been 
determined. We find significant changes in the coronal abundances 
during the flare with respect to the quiescent phase: all elements 
appear to be enhanced, except possibly Ne, and the Fe abundance in 
particular is found to increase by almost a factor 3.

Both abundances and EM(T) for \hr\ have been derived also by 
\cite{Nordon06}, and their findings are in agreement with the 
results presented here.

\subsection{Flaring loop modeling}
\label{s:res_modeling}
High level diagnostics of the flare can be obtained from detailed
hydrodynamic modeling of the flaring plasma \citep{Reale88,Reale04}.
The analogy of stellar to solar flares suggests that stellar as solar 
flares mainly occur in closed magnetic structures (coronal loops). 
It is customary to assume that the bulk of the flare involves a single
coronal loop, which can be then investigated through the analysis of 
the flare characteristics.

The plasma confined in coronal loops can be described as a compressible
fluid which moves and transports energy exclusively along the magnetic
field lines. We can then model a flare by solving the time-dependent
hydrodynamic equations in the coordinate along the loop with the
assumption of an appropriate transient input heating function.
The main model parameters include the loop length and the intensity,
distribution and duration of the heat pulse. These parameters are 
constrained by the observed flare evolution, in particular in the 
decay phase, and by parameters derived from the data analysis, such as
the temperature and emission measure at flare maximum and the timing 
of the maximum.
Scaling laws such as those shown in \citet{Reale97} help us to
constrain the loop length from the light curve decay time and the
flare maximum temperature:

\begin{equation}
L_9 = \frac{\tau_X \sqrt{T_7}}{120 f(\zeta)}
\label{eq:lscal}
\end{equation}
where $L_9$ is the loop half length ($10^9$ cm), $\tau_X$ the decay 
time of the light curve, $T_7$ the flare temperature ($10^7$~K), 
$f(\zeta) \ga 1$ a correction factor which takes into account possible 
significant heating during the decay, that might make the decay slower.

Other parameters, such as the loop initial atmosphere, or, for low 
gravity stars, the stellar surface gravity, have much less influence 
on the flare evolution.  We compute the evolution of the loop plasma 
by solving the time-dependent hydrodynamic equation of mass, momentum 
and energy conservation for a compressible plasma confined in the loop 
\citep{Peres82,Betta97}, including the relevant physical effects such 
as the plasma thermal conduction and radiative losses.
The gravity component along the loop is computed assuming a radius 
$R_* = 3 R_{\odot}$ and a surface gravity $g_* = 0.1 g_{\odot}$.

We integrate numerically the equations over a time range of 50~ks.
From the density and temperature distribution of the plasma along the
model loop, we synthesize the expected plasma X-ray model spectrum, 
and finally convolve with the \cha/{\sc meg} spectral response to produce 
synthetic counts spectra as described in \S\ref{s:analysis_hd}.

The flare light curve integrated in the whole \cha/{\sc meg} band resembles
quite closely the X-ray light curves of flares recently observed on PMS
stars \citep{Favata05}. For this reason, we have assumed as a first
set of parameters the same values which best describe a specific flare
observed in the Orion region, i.e.\  a loop with constant cross-section
and half-length $L = 10^{12}$~cm, symmetric around the loop apex. 
The flare simulation is triggered by injecting a heat pulse in the loop
which is initially at a temperature of $\sim 20$ MK. 
Two heat pulses are deposited with a Gaussian spatial distribution of 
intensity 10~erg~cm$^{-3}$~s$^{-1}$ and width $10^{10}$~cm (1/100 of the 
loop half-length) at a distance of $2 \times 10^{10}$~cm from the 
footpoints, i.e.\ very close to them \citep{Reale04}. 
After 20~ks the heat pulses are switched off completely. 
With the above parameters, the light curve is reproduced with good 
accuracy but we have noticed that the temperature values obtained from 
fitting the spectra with isothermal plasma emission models are 
significantly higher than those derived from the data all along the 
flare evolution \citep{Testa05b}.
Therefore, we have refined the model parameters to improve the fitting:
we have reduced the input heating enough to account for the desired
temperature decrease.  From Eq.(\ref{eq:lscal}), in order to maintain 
the same decay time, a temperature reduction implies a shorter loop 
length.  After this feedback, we have obtained best results with a loop 
half-length $L = 5 \times 10^{11}$~cm (half as before) and a heat pulse 
of 4~erg~cm$^{-3}$~s$^{-1}$  (peak volumetric heating) lasting 15~ks 
(5~ks less than before). 
The total energy flux rate is $2.0 \times 10^{11}$~erg~cm$^{-2}$~s$^{-1}$.

\begin{figure}[!h]
\centerline{\psfig{figure=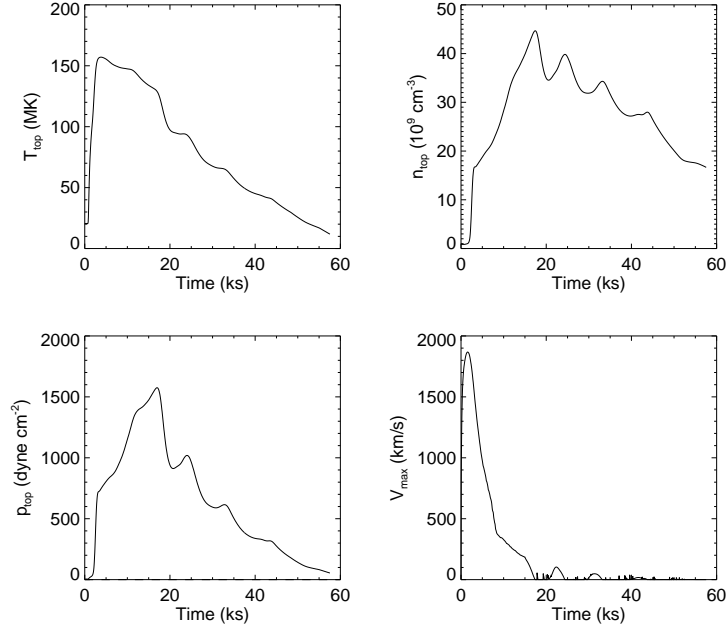,width=10cm}}
\vspace{-0.3cm}
\caption{Evolution of the temperature, density and pressure at the loop 
	apex and of the maximum plasma velocity according to the loop
	hydrodynamic model of the flare.
	\label{fig:apex}}
\end{figure}

The computed evolution largely resembles the evolution computed for
other stellar flares, although on larger scales than for typical
flares observed in late-type stars coronae (e.g.\ 
\citealt{Reale04}).  
Fig.~\ref{fig:apex} shows the evolution of the temperature, density, 
pressure at the apex of the loop and of the maximum plasma velocity.
The heat pulses make the loop plasma heat rapidly ($\sim 3$~ks) 
at temperatures about 150~MK and expand dynamically upward from
the chromosphere at speeds about 1800~km~s$^{-1}$ to reach the loop 
apex also in about 3~ks. 
After this first impulsive phase, the temperature slowly decreases to 
about 130~MK while the heat pulse is on, and the evaporation of plasma
from the chromosphere -- which brings the density increase shown in 
Fig.~\ref{fig:apex} -- continues substantially but less dynamically, 
with plasma speeds below 500~km~s$^{-1}$ after 9~ks, when the plasma 
pressure becomes higher than 1000~dyne~cm$^{-2}$.
The plasma pressure and density reach their peak of 1500~dyne~cm$^{-2}$
and $4.5 \times 10^{10}$~cm$^{-3}$ slightly (about 3~ks) later than 
the end of the heat pulse. Then they begin to decrease gradually
with significant quasi-periodic fluctuations, while the temperature 
initially drops and then decreases more gradually. 

\begin{figure}[!h]
\centerline{\psfig{figure=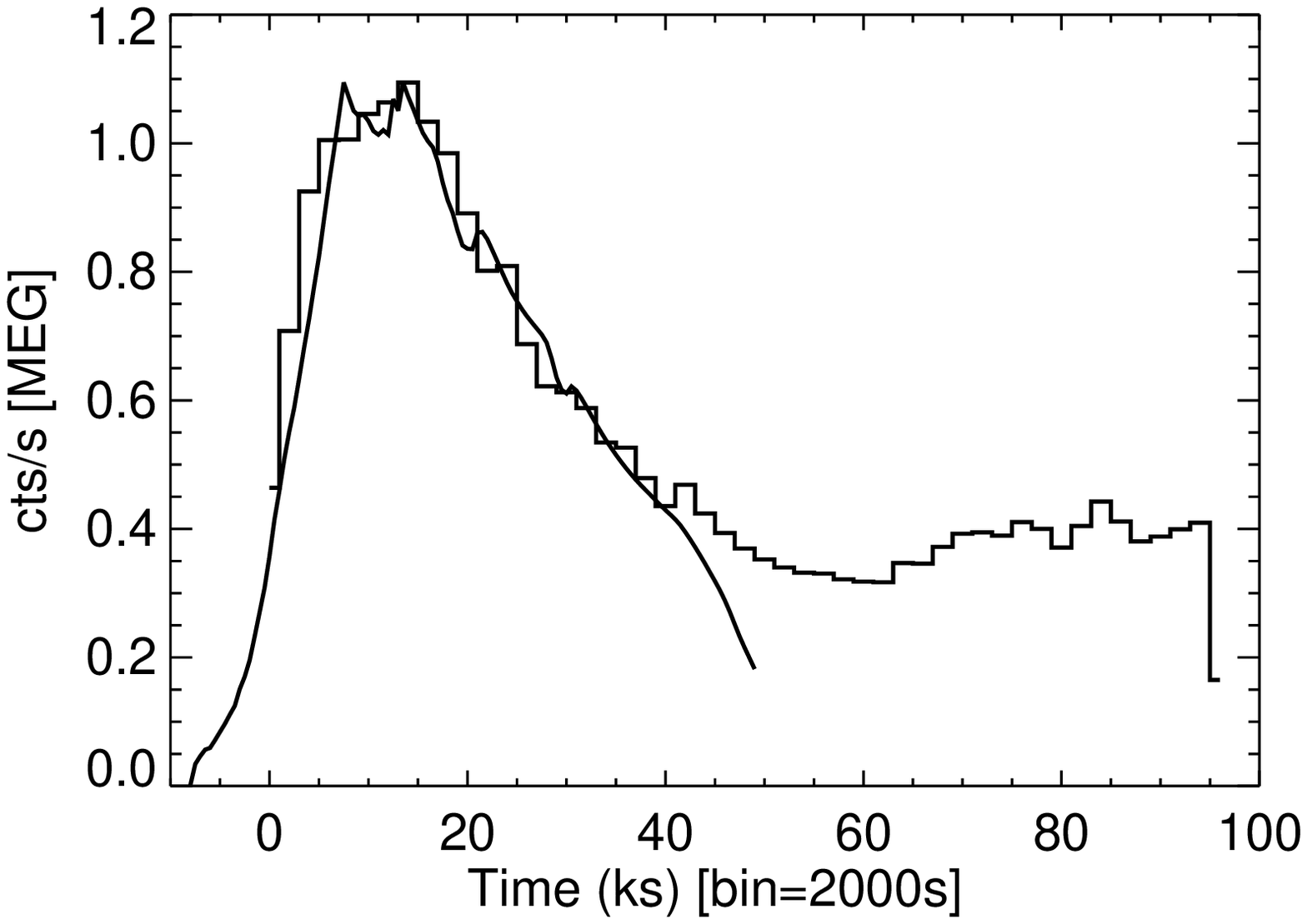,width=9cm}\hspace{-0.6cm}
	\psfig{figure=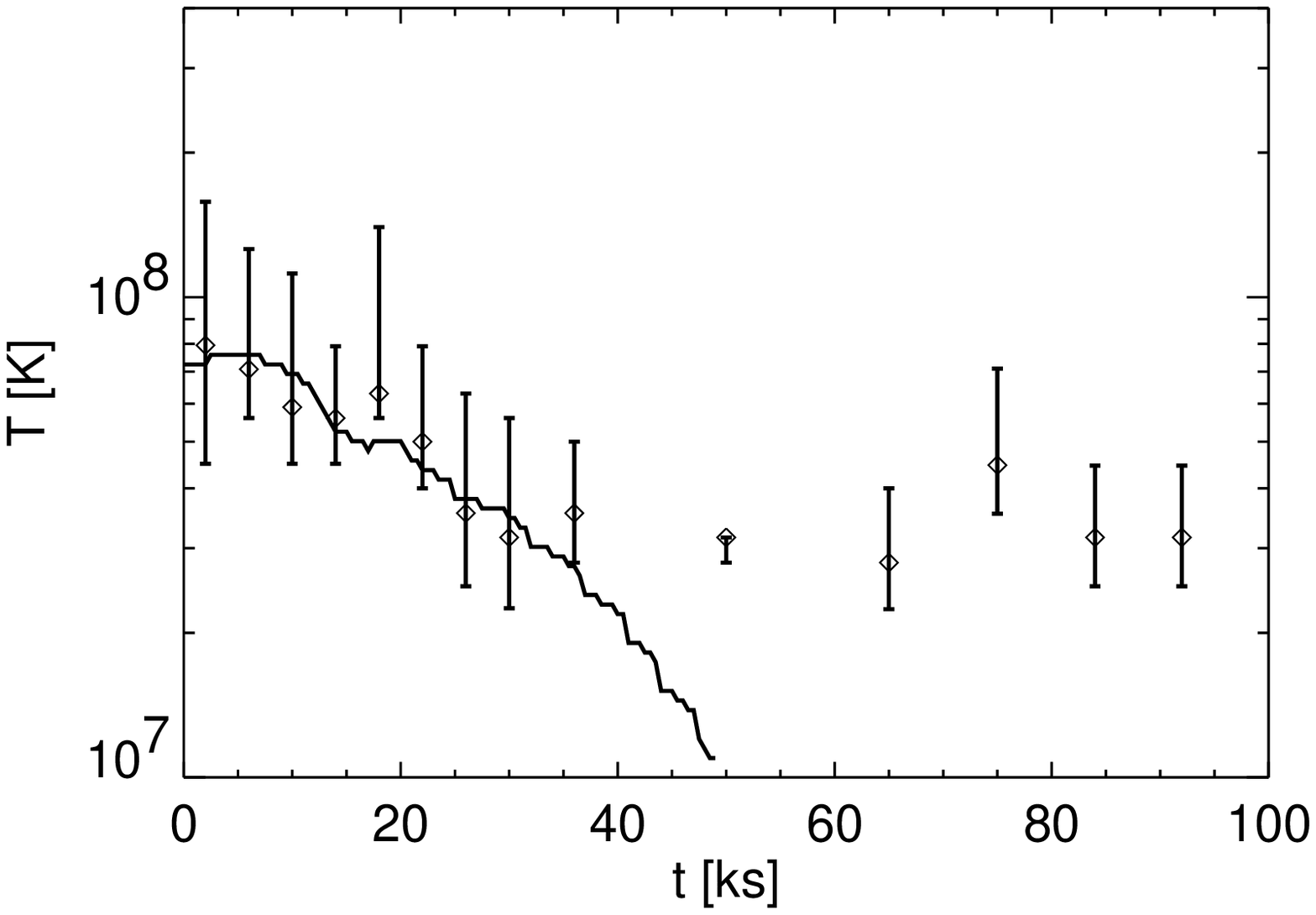,width=9cm}}\vspace{-0.5cm}
\centerline{\psfig{figure=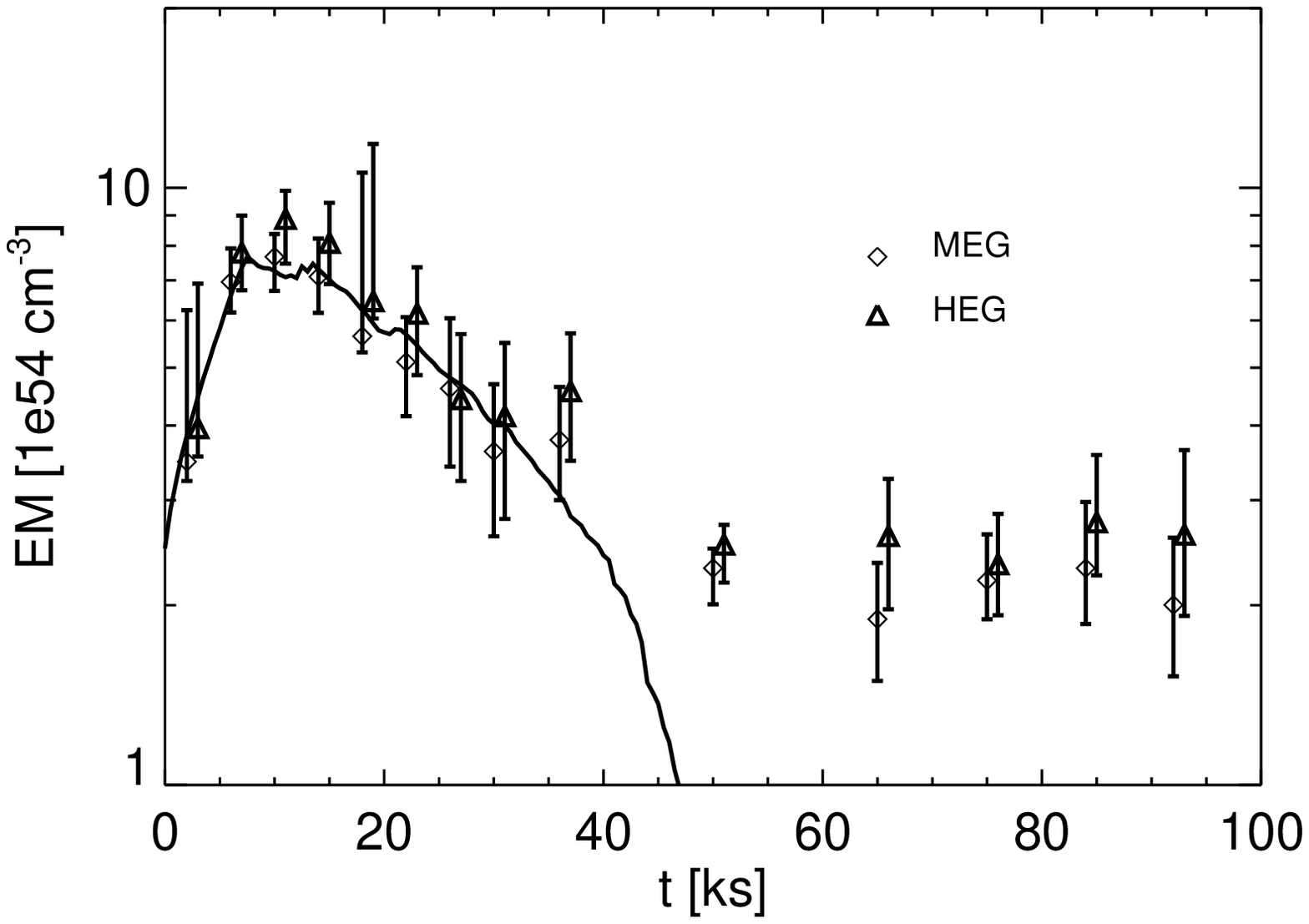,width=9cm}\hspace{-0.6cm}
	\psfig{figure=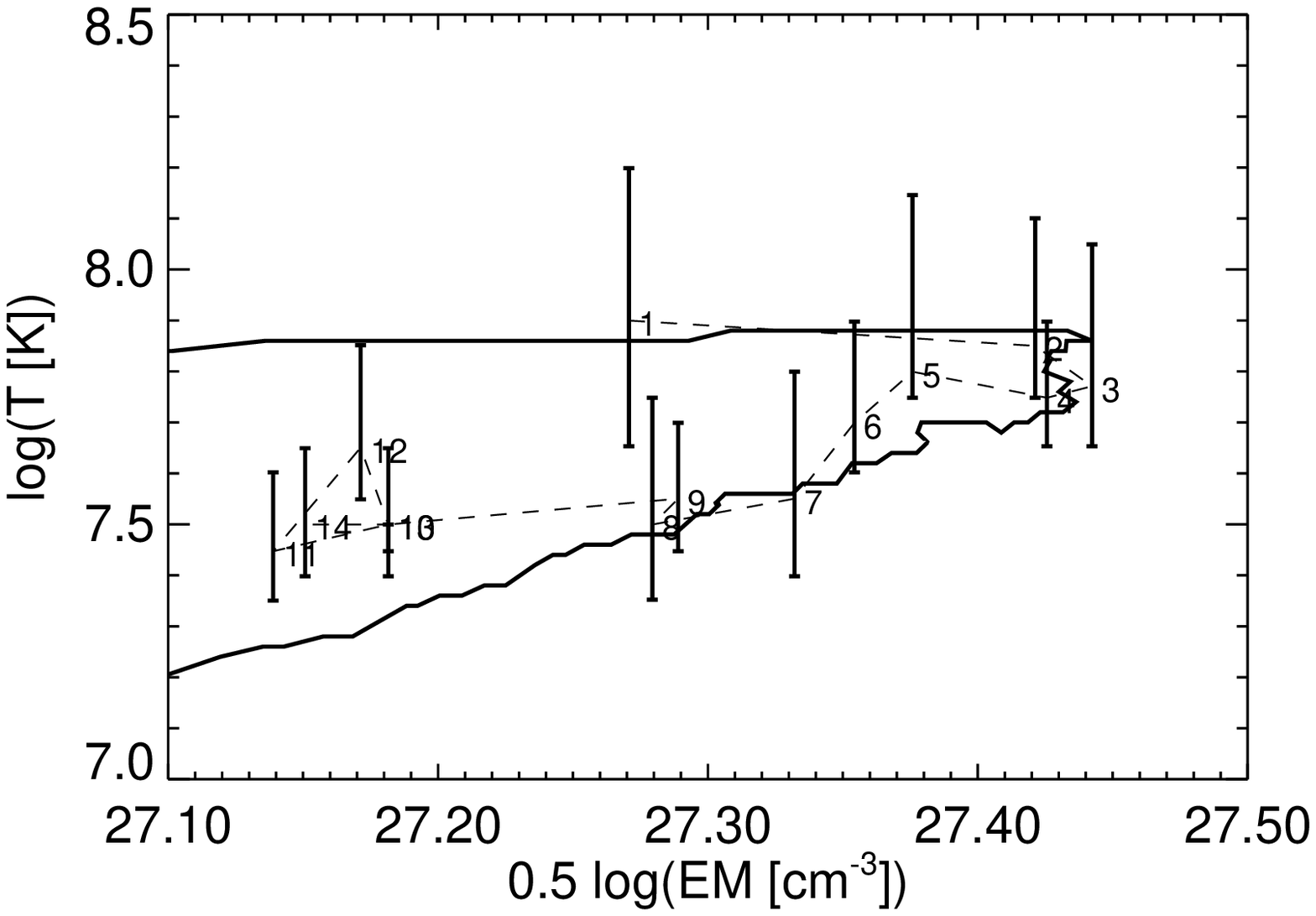,width=9cm}}\vspace{-0.5cm}
\caption{Comparison of observed lightcurve (top left), T (top right),
	EM (bottom left), and T-n (bottom right) evolution, with the 
	corresponding quantities synthesized from the hydrodynamic
	loop model (solid thick lines) characterized by half-length
	$L = 5 \times 10^{11}$~cm, and footpoint impulsive 
	heating. Evolution of temperature and emission
	measure (top right, and bottom left panels) are derived from
	the fit with isothermal models to the continuum emission in
	the spectra integrated in the time intervals shown in 
	Fig.~\ref{fig:lightcurve}. 
	\label{fig:modelfit}}
\end{figure}

Figure~\ref{fig:modelfit} shows the integrated light curve 
assuming a loop cross-section radius of $4.3 \times 10^{10}$~cm,
i.e. 8\% of the loop half-length. The spectra obtained from the loop
modeling have also been fit with isothermal models. The resulting
evolution of the emission measure and of the temperature, and the T 
vs.\ EM diagram are shown in the figure. A visual comparison of the 
model results with the data and the data fitting results indicates 
that the loop simulation results are in good agreement with the data.

The dynamic model described above closely matches the evolution of 
temperature and emission measure derived from the continuum emission, 
and the lightcurve integrated over the entire wavelength range. 
The high quality data allows us to test the model and perform a 
detailed comparison of its characteristics with the observations.
In Figure~\ref{fig:spec_mod_data} we compare the {\sc meg} spectra 
synthesized from the model with the observed spectra in 4 different
phases of the flare: rise (0-10~ks from beginning of observation), 
peak (10-15~ks), early decay (15-30~ks), and late decay (30-45~ks). 
Figure~\ref{fig:spec_mod_data} shows a general good agreement of the
model spectra with the observations; however, some systematic 
discrepancies are observed. For example, in the spectrum of the late 
decay phase several lines, such as the \sixiv\ \lya\ ($\sim 6.2$\AA), 
the hot Fe lines around 11\AA, and the \oviii\ \lya\ ($\sim 19\AA$),
are overpredicted.
There are several factors that may determine these departures from the
observations. For instance, the abundances used for synthesizing the 
model spectrum are the ones derived from the flare spectrum and are 
kept fixed.  Since we observe significant changes in the abundances 
after the flare (\S\ref{s:res_demabund}) they may change on short time 
scales and affect the emission in the spectral lines. Another process, 
not taken into account in our modeling, that might affect the emission 
in the lines on short time scale is non-equilibrium ionization; 
however, this effect is expected to be relevant only at the very 
beginning of the flare (see e.g., \citealt{Orlando03b}).  
Furthermore, in our simulation we assume that all the observed 
emission comes from the flaring structure, i.e.\ the emission of the 
quiescent corona (or of other secondary flaring structures) is assumed 
to be negligible whereas it could in fact contribute significantly to 
the ''cool" ($\log T[$K$] \lesssim 7.2$) lines at least, as suggested 
by the derived emission measures during flare and quiescence 
(Fig.~\ref{fig:dem}).

\begin{figure}[!ht]
\centerline{\psfig{figure=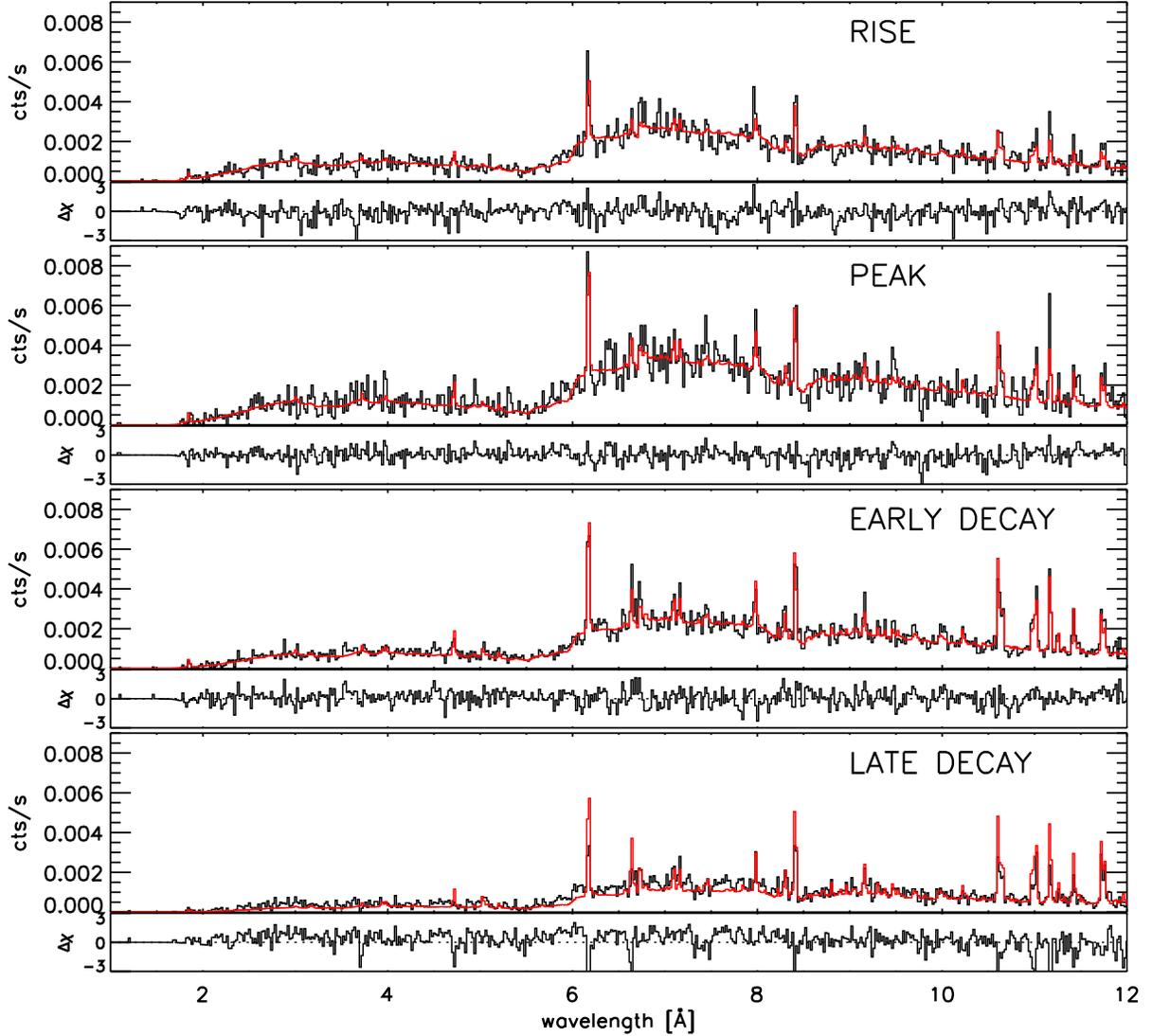,width=16cm}}
\vspace{-0.3cm}
\caption{Comparison of \hr\ {\sc meg} spectrum (black histogram) and 
	the {\sc meg} spectrum synthesized from the hydrodynamic model 
	(red histogram) in four different phases of the flare, from 
	top to bottom,: rise (0-10~ks from beginning of observation), 
	peak (10-15~ks), early decay (15-30~ks), and late decay 
	(30-45~ks). Spectra are shown in counts per second, and for 
	better readability we split the spectral range in two plots
	showing the spectral range 1.5-12\AA\ and 12-22\AA\ respectively.
	The small panel below each plot shows the $\chi$ residuals.
	\label{fig:spec_mod_data}}
\end{figure}
\addtocounter{figure}{-1}
\begin{figure}[!ht]
\centerline{\psfig{figure=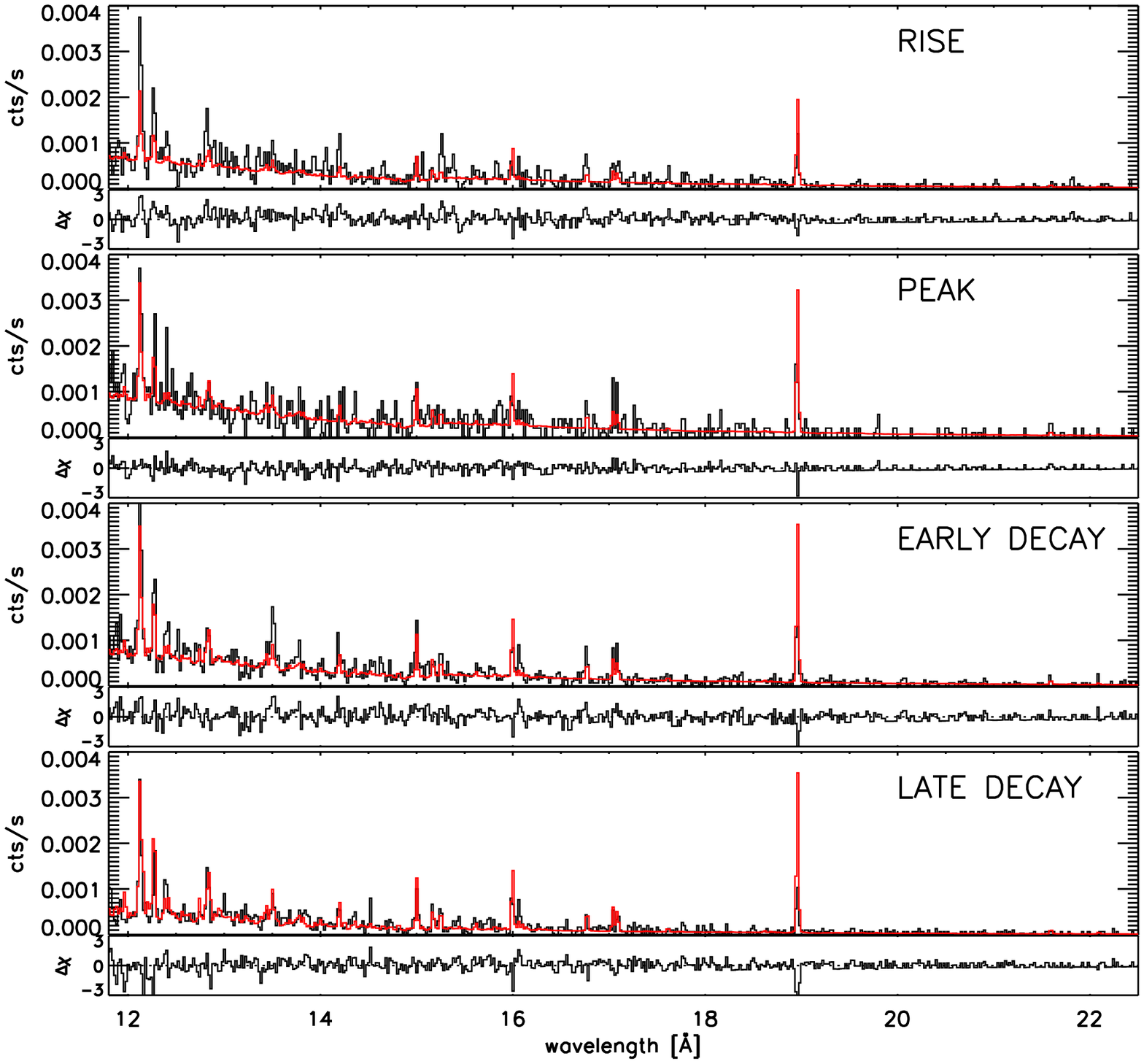,width=16cm}}
\vspace{-0.3cm}
\caption{continued.}
\end{figure}

\begin{figure}[!h]
\centerline{\psfig{figure=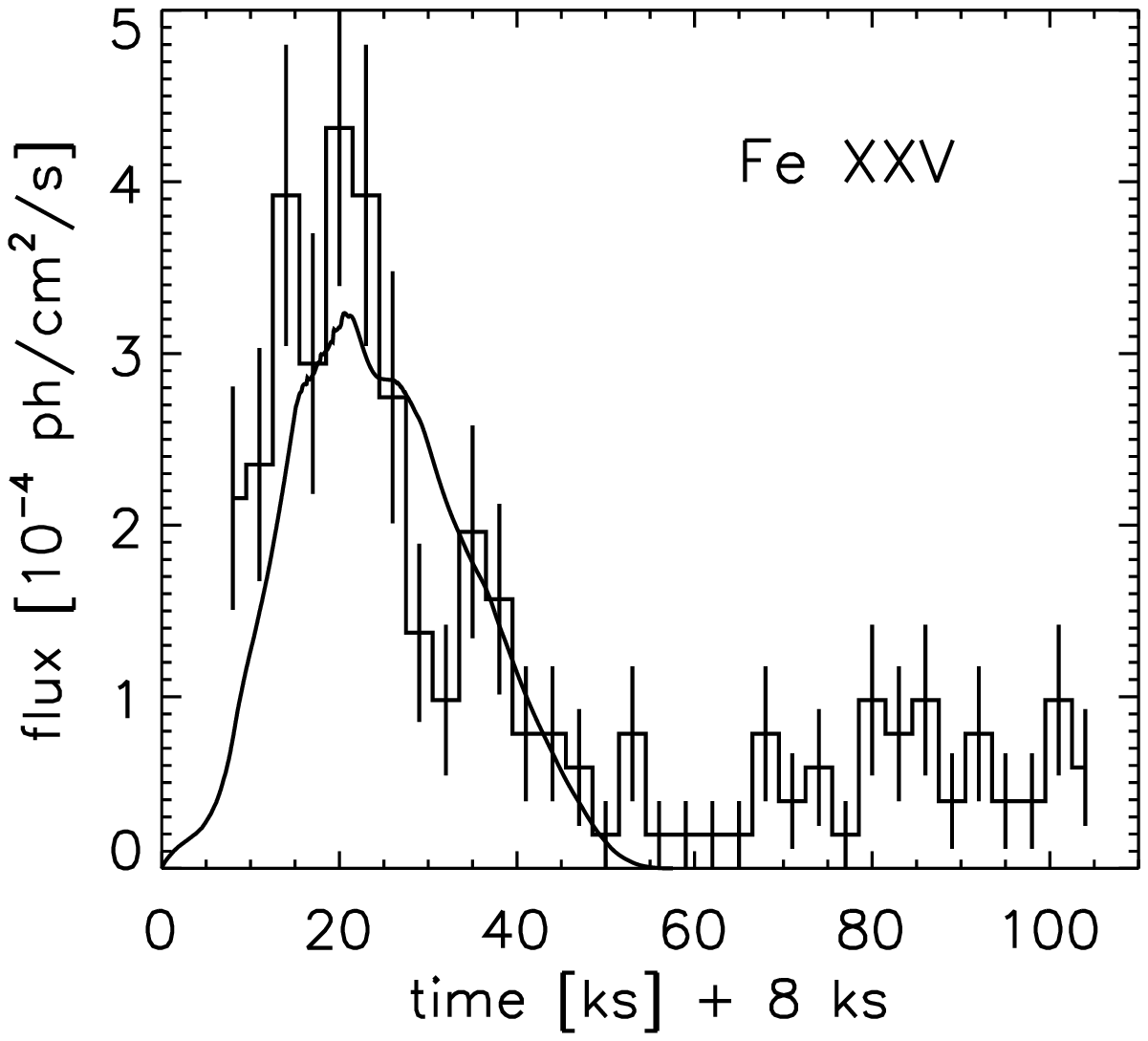,width=7cm}\hspace{-0.7cm}
	    \psfig{figure=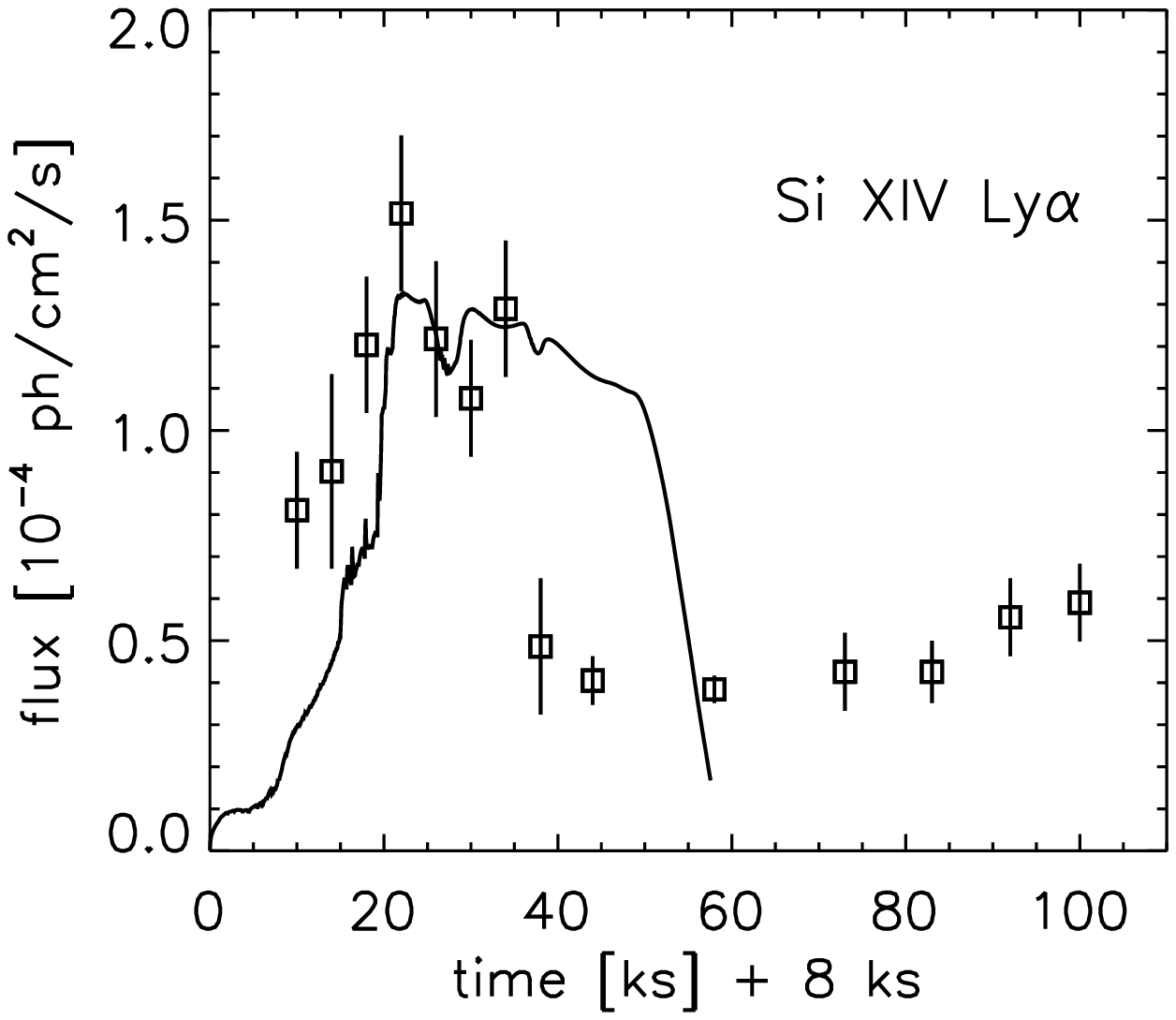,width=7cm}\hspace{-0.7cm}
	    \psfig{figure=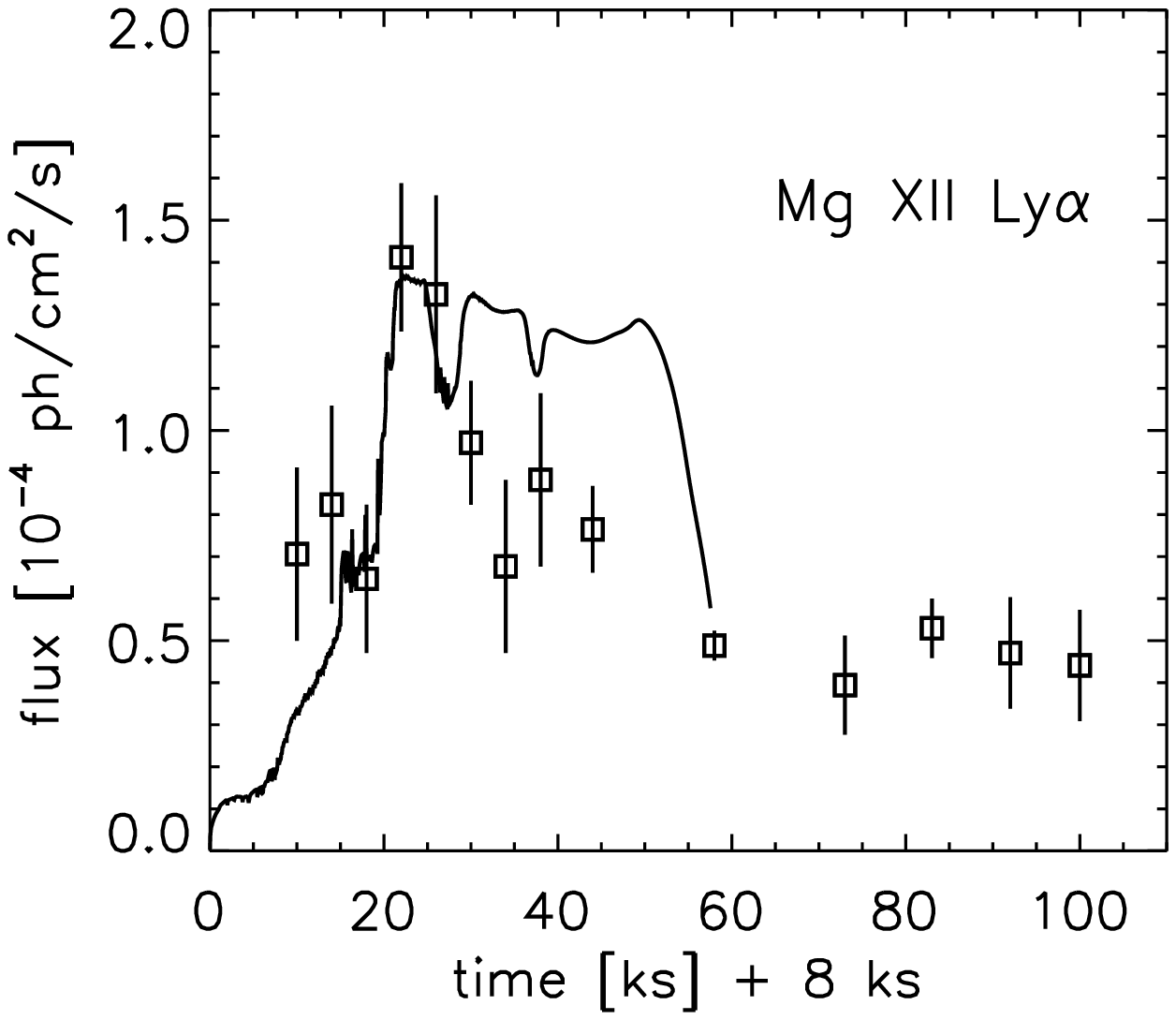,width=7cm}}
\vspace{-0.3cm}
\caption{Lightcurves derived from \hetg\ spectra in selected spectral 
	features (histograms with error bars) compared with the 
	prediction from the hydrodynamic model (thick solid line),
	assuming a loop cross-section radius $4.8 \times 10^{10}$~cm.
	\label{fig:lclin_mod_dat}}
\end{figure}

The high quality of the data allows us to test the model to a much 
higher level of detail than before, for instance allowing a direct 
comparison with the lightcurves in single spectral features.
The evolution in very hot spectral features emitted at the bulk plasma
temperature of the flaring loop (i.e.\ $\log T[$K$] > 7.5$) would 
represent more meaningful tests for the model. However, there are only 
a few very hot lines present in the spectrum (e.g., \fexxv\ and \caxx), 
and only the \fexxv\ complex has enough signal to provide temporal 
resolution.
Beside \fexxv, we selected two other emission lines with high enough 
flux, \sixiv\ and \mgxii\ \lya\ lines, and derived their lightcurves 
using the same time intervals used for the analysis of the continuum 
and marked in Fig.~\ref{fig:lightcurve}.
Figure~\ref{fig:lclin_mod_dat} compares the lightcurves in these 
spectral features derived from the observations with those synthesized
from the hydrodynamic simulation, assuming a loop cross-section radius 
$r = 4.8 \times 10^{10}$~cm.
The line fluxes for \sixiv\ and \mgxii\ are obtained from the fit to 
the spectra integrated in each time interval. For the \fexxv\ complex 
the lower signal to noise ratio does not constrain the fit; however, 
since in the relevant wavelength range the continuum emission is 
rather small with respect to the line emission we use the lightcurve 
integrated in a small wavelength range (1.83-1.89\AA) using a 3~ks 
temporal bin.

The model reproduces extremely well the observed temporal behaviour 
of the \fexxv\ line complex. The lightcurves of \sixiv\ and \mgxii\ 
derived from the observation and from the model lie close to each 
other up to the peak of the flare. After the peak the observed 
emission in the two lines experiences the dramatic drop predicted from
the model, but on much shorter timescales, anticipating the model by 
slightly more than 10~ks.  Other than the effects discussed above 
possibly causing this discrepancy, we can identify another effect 
that may play a role.   
In the model of the flaring loop the plasma contributing the most to 
these ''cool" lines ($\log T[$K$] \sim 7.0-7.2$) is located in the 
transition region of the loop, that is the region where the plasma 
temperature and density steeply change from the dense and cool 
chromosphere to the hot and rarefied coronal portion of the loop. 
In this region some models predict an opening of the magnetic flux 
tube which could have a substantially larger cross-section in corona 
than in the low region connecting it to the photosphere (e.g., 
\citealt{Gabriel76,Schrijver89,Litwin93,Ciaravella96}).
In this possible scenario, the emission of the flaring structure in 
these cool lines would be substantially reduced and the resulting 
excess would be therefore attributed to the background corona or to 
nearby secondary flaring structures (e.g., \citealt{Betta01}).

\begin{figure}[!h]
\centerline{\psfig{figure=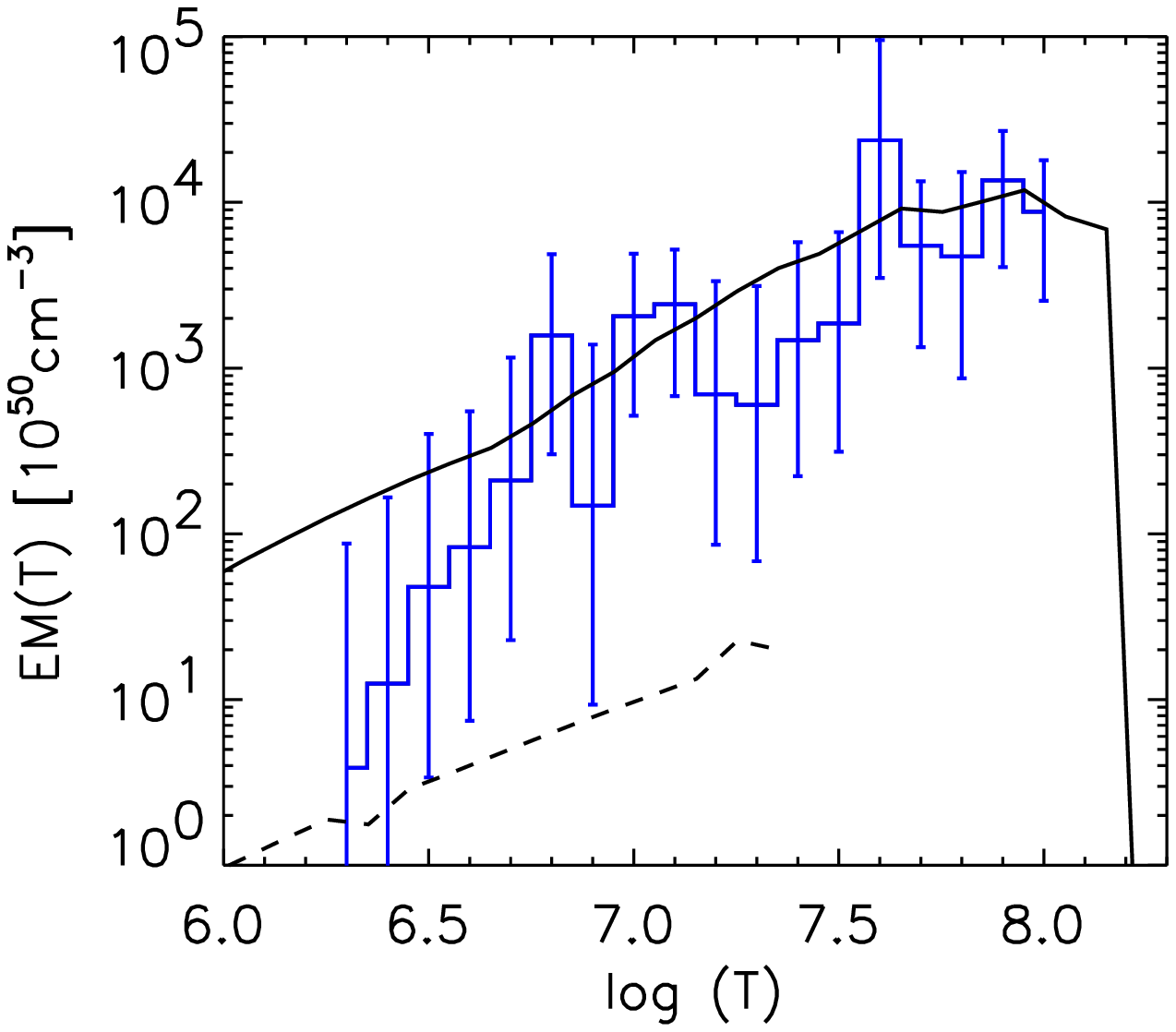,width=10cm}}
\vspace{-0.3cm}
\caption{Comparison of emission measure distribution derived from the 
	flare spectrum (histogram with error bars) with the EM(T) of 
	the hydrodynamic loop model (solid thick line). The dashed 
	line shows the EM(T) of the initial static loop with maximum 
	temperature of 20~MK.
	\label{fig:em_dat_mod}}
\end{figure}

Finally, we compared the emission measure distribution of the loop 
model with the EM(T) derived from the flare spectrum. 
Figure~\ref{fig:em_dat_mod} shows that also for the emission measure 
distribution the model predictions are in very good agreement with 
the observations. This agreement is a further indication of the
validity of the single loop model for this flare and, on the other 
hand, of the EM inversion method.
In our modeling of the X-ray emission during the flare we 
assume any persistent quiescent emission to be negligible. 
As mentioned in \S\ref{s:res_demabund}, the \cha\ observation does 
not provide a good determination of the quiescent emission as outside 
the large flare constituting the focus of our analysis the corona is 
undergoing another significant, though smaller, dynamic event 
(its average \lx\ $\sim 4.2 \times 10^{31}$~erg~s$^{-1}$ is about 
twice the luminosity value of other X-ray observations; see 
\S\ref{s:intro} and \S\ref{s:discuss}). Other
X-ray observations of this corona suggest that the quiescent
emission is in fact negligible to a large extent with respect to
the large and hot flare observed with \cha\ (see \S\ref{s:discuss}). 

\section{Discussion and conclusions}
\label{s:discuss}

Our analysis has shown that the observed evolution of the very hot 
coronal emission of the single giant \hr\ is reproduced extremely well
by a model characterized by loop semi-length $L = 5 \times 10^{11}$~cm 
($\sim R_{\star}/2$), and impulsive footpoint heating triggering the 
flare.   The heating pulse lasts 15~ks, and it is shifted by 8~ks 
preceding the beginning of observation.  The peak volumetric heating 
is $4$~erg~cm$^{-3}$~s$^{-1}$, with a corresponding total heating rate 
$\approx 10^{33}$~erg~s$^{-1}$. The detailed analysis of the plasma 
evolution in the temperature-density diagram (Fig.~\ref{fig:modelfit}) 
put tight constraints on the flare morphology and on the characteristics 
of the heating. The delay of the EM peak with respect to the temperature 
peak indicates that the X-ray emitting plasma is confined in a closed 
structure, whereas in non-confined coronal flares the EM evolves 
simultaneously with the temperature as shown by the analysis of 
hydrodynamic models \citep{RealeBP02}.
The steep slope of the decay path in the temperature-density diagram
implies that after the initial phase when the energy is released, the 
loop undergoes a pure cooling evolution, and no addition heating is 
needed to explain the evolution of the flaring plasma.

As discussed in \S\ref{s:res_modeling} the initial phases of the flare
are very dynamic and the plasma fills up the loop expanding upward at 
very high speed reaching $\sim 1800$~km~s$^{-1}$. After about 
$\sim 10$~ks, therefore corresponding to the start of the \cha\ 
observation (considering the shift of 8~ks between the energy release 
and the start of the observation), the plasma speed rapidly decreases; 
however, during the first 5-10~ks of the observation plasma flows with
speed up to a few 100~km~s$^{-1}$ are present.  Such speed can be in 
principle resolved by \hetg.  The plasma characterized by such high 
speed is plasma at temperatures above 60~MK, therefore, in order to 
investigate these speeds through Doppler shifts we have to search for 
shifts in very hot lines. However, such hot lines, e.g.\ \caxx, are at
low $\lambda$ where the effective area is low and the instrument is 
not very sensitive. 
We searched for shifts in the \caxx\ line but we did not detect any 
significant shift.  We note that, even in the case the instrument 
sensitivity were not a limiting factor, the detection of line shifts 
is expected only for a preferential orientation of the loop with 
respect to the line of sight.

The high photon statistics of the \hetg\ data allowed us to derive
light curves in a few relevant hot lines and, for the first time for
a stellar flare, they could be compared in detail to those predicted 
by a hydrodynamic model (see \citealt{Peres87,Betta01}, for analogous 
comparisons for a solar flare).

To a more general level, we could also compare the distributions of 
emission measure obtained from the data analysis and from the flare
loop model and found a good agreement. The shape of the EM 
distribution is quite typical of a single coronal loop (e.g., 
\citealt{Peres00}) and different from those found for active stars 
(e.g., \citealt{Sanz02,Testa05a,Cargill06}).

\begin{figure}[!h]
\centerline{\psfig{figure=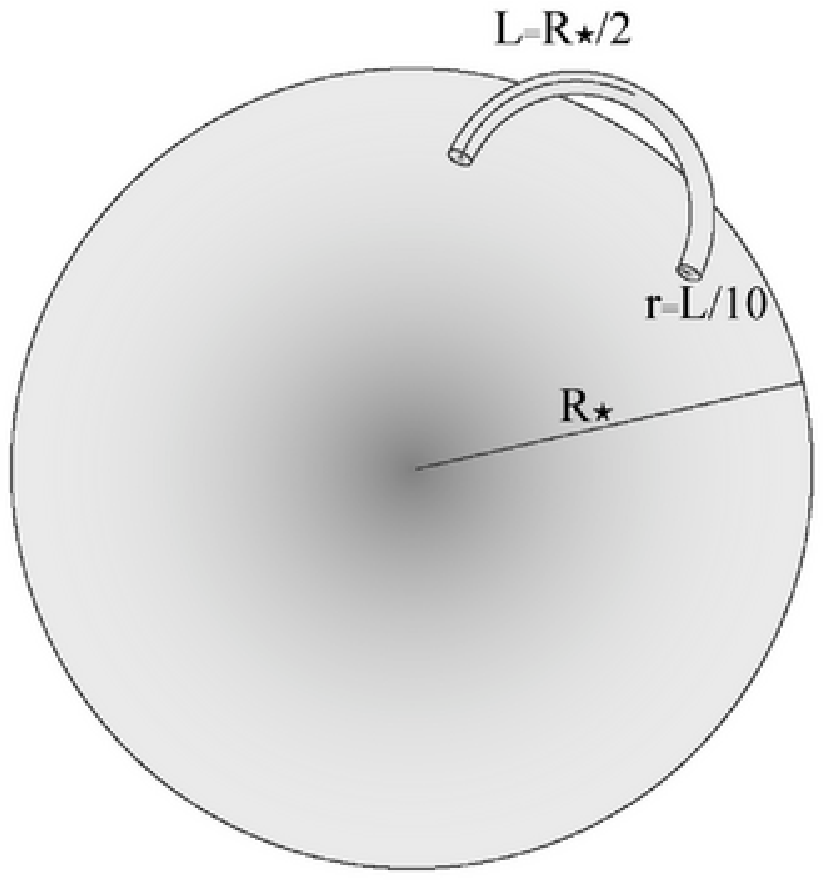,width=7cm}}\vspace{-0.3cm}
\caption{Sketch showing the geometry of the flaring loop of \hr\ as 
	derived from the hydrodynamic modeling of the flare: R$_\star$ 
	is the stellar radius, L is the loop semi-length, and r is the 
	cross-section radius of the loop.
	\label{fig:sketch}}
\end{figure}

From the normalization of the model lightcurves we derive an estimate 
of the loop cross-section and therefore of its aspect ratio 
$\alpha=r/L$.  We note that the loop cross-section radius, $r$, is a 
free parameter for which we can obtain independent estimates from the 
normalization of the different lightcurves: (a) integrated {\sc meg} 
dispersed counts, (b) EM derived from the analysis of the continuum 
(Fig.~\ref{fig:modelfit}), and (c) fluxes of spectral features 
(Fig.~\ref{fig:lclin_mod_dat}), therefore providing a cross-check
of the consistency of our model.            
The obtained values of $r = 4.9 \times 10^{10}$~cm, 
$4.3 \times 10^{10}$~cm, and $4.8 \times 10^{10}$~cm respectively, all 
agree within 15\%, and they imply a loop aspect ratio $\alpha \sim 0.09$.
Figure~\ref{fig:sketch} shows a sketch of the morphology of the flaring 
structure as inferred from our modeling of the observed \cha-\hetg\ 
spectra.

As mentioned in \S\ref{s:res_modeling}, the loop model reproducing the
observation has roughly the same parameters of models satisfyingly 
reproducing other large flares, specifically the flares observed in 
pre-main sequence stars \citep{Favata05}.  However, the derived loop 
parameters reveal some dissimilarity possibly indicating fundamental 
differences between the processes at work in the two cases.  The 
flaring loop reproducing the flare on \hr\ is characterized by loop 
semi-length $L \sim R_{\star}/2$, and aspect ratio $r/L \sim 0.1$, 
as typically found for low mass stars and on the Sun. In contrast, 
the modeling of the flare on the pre-main sequence stars, in particular 
the detailed model for COUP source 1343, yields a very large loop 
semi-length corresponding to several stellar radii, and with a much 
smaller aspect ratio $r/L \sim 0.02$ \citep{Favata05}.
These findings suggest that \hr\ is characterized by a ''normal" 
corona, while in the young stars showing evidence of elongated X-ray 
emitting structures, possibly connecting the star to the disk, 
processes profoundly different may be at work.

Large flares as the one observed in \hr\ are very unusual in single 
intermediate mass giants, and there are only a few such events studied 
in detail.  This study thus allowed us to investigate the flaring 
activity in a single giant, and provided us with a powerful tool to 
derive the size of its coronal structures.  Even though the flaring 
structure modeled here is not necessarily representative of the 
typical coronal structures in X-ray active giants, as discussed above,
its characteristics are consistent with the presence of a corona with 
properties typical of active coronae but scaled up to the larger 
stellar radius.
The coronal properties derived through X-ray observations of these 
active giants might suggest a possible interpretation for the low 
frequency of flares.  There is some evidence of very low coronal 
surface filling factors ($\lesssim 10^{-4}$) for these active giants 
as suggested from density analysis (e.g., \citealt{Testa04b}).
Considering the interaction of magnetic fields in active regions as a 
possible mechanism increasing the X-ray flaring activity in active 
stars (\citealt{Guedel97,Testa04b}), in these stars this kind of 
interaction might be much less frequent given to the very sparse 
presence of active region distributed on a large area.

In this paper both for the modeling and its interpretation we have 
assumed that the quiescent emission is negligible with respect to the
flaring emission. The \cha\ observation does not provide a good grasp 
of the quiescent emission as the hardness ratio behaviour (and the 
emission measure distribution) suggests the presence of another
dynamic event subsequent to the large flare (as in \citealt{Reale04}). 
The \xmm\ observations might represent the quiescent X-ray spectrum of 
\hr\ not showing significant variability, however their very low 
exposure time (6~ks, 3~ks) allow us only to assess the variability on 
very short time scales. Our preliminary results of a $\sim 50$~ks new 
Suzaku observation of \hr\ shows extremely constant X-ray emission and 
provides an estimate of the quiescent X-ray luminosity and temperature 
of about $2 \times 10^{31}$~erg~s$^{-1}$ and $3 \times 10^7$~K 
respectively (P.Testa et al.\ in prep.), consistent with the findings 
based on \xmm\ and ROSAT observations \citep{Gondoin03,Singh96a}.
This X-ray luminosity value which we assume being a good estimate for
the quiescent emission of \hr, is about an order of magnitude lower 
than the peak luminosity of the flare observed with \cha, therefore 
lending support to the assumption that the quiescent emission is
negligible with respect to the flare emission.  Also, the main 
results of our study rely on the emission of the very hot plasma which
is completely dominated by the flaring structure, and therefore they
should not be significantly affected by ignoring the much cooler
quiescent emission.

In this work we model the flare emission with a single loop 
structure and it is worth discussing whether and to what extent very
different solutions can be ruled out.
We model the bulk of the flare where presumably there is one dominant 
loop structure, as observed in many solar compact flares entirely 
occurring in single loops and even in more complex flares, where one 
can often consider a dominant loop (e.g.\ \citealt{Aschwanden01}).
The single loop model satisfactorily reproduces the \cha\ observations 
analyzed in this work and multiple loop models would not add insight 
and would include more free and unconstrained parameters. 
Furthermore, the single loop model is supported by the fact that : 
(1) the loop cross-section area is consistent with that of a single 
loop structure; (2) the thermal distribution is compatible with 
that of a single flaring loop; (3) the decay path in the 
density-temperature diagram 
has quite a steep slope ($\sim 2$) implying negligible heating in the 
decay \citep{Sylwester93}, in turn indicating a single loop structure, 
whereas flaring arcades are characterized by significant heating 
(e.g.\ \citealt{KoppPoletto}).
At later times of our observation (when the light curve rises again) 
other loops may be involved and become important in the evolution 
(e.g.\ \citealt{Reale04}).  
The general expression of the loop length as a function of the observed
decay time (Eq.~\ref{eq:lscal}) also allows us to estimate the uncertainties
on the derived loop length. In particular, using the expression of the 
correction factor $f(\zeta)$ tuned for Chandra, as reported in 
\cite{Favata05}, we obtain $L_9 = 490$. Considering typical uncertainties
of the diagnostic formula, and the uncertainties on the observed 
temperature, we estimate an error of about 20-30\% on the loop length,
therefore yielding $L = 4.9 \pm 1.5 \times 10^{11}$~cm. It is worth
noting that this formula used for estimating the uncertainties on the 
derived loop geometry assumes uniform heating and provides only 
approximate values for the loop length. The 
hydrodynamic model provides us with a much higher level of detail,
giving us diagnostics for the temporal and spatial distribution of 
the heating, for the loop aspect, and for the thermal distribution of 
the plasma.

\begin{acknowledgements}
PT and DPH were supported by SAO contract SV3-73016 to MIT for support of CXC, 
which is operated by SAO for and on behalf of NASA under contract NAS8-03060.
FR acknowledges support from Agenzia Spaziale Italiana and italian Ministero 
dell'Universit\`a e della Ricerca.
\end{acknowledgements}

\end{document}